\let\orilabel\label
\documentclass[aps,twocolumn]{revtex4-2}
\let\label\orilabel

\usepackage{amsfonts,amsmath,bm,amssymb,revsymb,color,braket}
\usepackage{xcolor}
\usepackage{graphicx}
\usepackage[makeroom]{cancel}
\definecolor{bostonuniversityred}{rgb}{0.8, 0.0, 0.0}
\definecolor{dukeblue}{rgb}{0.0, 0.0, 0.61}
\definecolor{ao(english)}{rgb}{0.0, 0.5, 0.0}
\definecolor{darkmagenta}{rgb}{0.55, 0.0, 0.55}
\definecolor{armygreen}{rgb}{0.29, 0.33, 0.13}
\definecolor{coquelicot}{rgb}{1.0, 0.22, 0.0}
\definecolor{fucsiak}{rgb}{0.4, 0.08, 0.4}
\definecolor{airforceblue}{rgb}{0.36, 0.54, 0.66}
\definecolor{applegreen}{rgb}{0.55, 0.71, 0.0}
\definecolor{awesome}{rgb}{1.0, 0.13, 0.32}
\definecolor{burgundy}{rgb}{0.5, 0.0, 0.13}
\definecolor{cobalt}{rgb}{0.0, 0.28, 0.67}
\definecolor{aqua}{rgb}{0.0, 1.0, 1.0}
\definecolor{blue-green}{rgb}{0.0, 0.87, 0.87}

\usepackage[unicode=true,bookmarksnumbered,colorlinks=true, citecolor=burgundy, linkcolor=blue, urlcolor=cobalt]{hyperref}

\def\ky{k_y}
\def\oky{\overline k_y}
\renewcommand{\Re}{\mathrm{Re}}

\begin{document}

\title{Absorption and reflection of inertial waves by a geostrophic vortex}

\author{Nikolay A. Ivchenko}

\affiliation{Landau Institute for Theoretical Physics,
Russian Academy of Sciences,\\1-A Akademika Semenova av.,
142432 Chernogolovka, Russia}

\affiliation{National Research University Higher School of
Economics, Laboratory for Condensed Matter Physics, 101000 Moscow, Russia}

\email{ivchenko@itp.ac.ru}

\author{Sergey S. Vergeles}

\affiliation{Landau Institute for Theoretical Physics,
Russian Academy of Sciences,\\1-A Akademika Semenova av.,
142432 Chernogolovka, Russia}

\affiliation{National Research University Higher School of
Economics, Faculty of Physics, Myasnitskaya 20, 101000
Moscow, Russia}

\email{ssver@itp.ac.ru}

\begin{abstract}
We study interaction of inertial waves with geostrophic flow in a rapidly rotating fluid system. In~accordance with experimental conditions in \cite{tumachev2023tworegimes,tumachev2024observation}, we consider inertial waves, which were excited by a source being near the side boundary of the flow and enter the region where geostrophic vortex flow is present. The wave equation is derived and analyzed in the paper that describes the propagation of convergent and divergent cylindrical waves on the background of mean vortex flow, which is considered as an axisymmetric differential rotation. We show that a monochromatic wave does not exert any torque on the vortex flow in the inviscid limit until it is absorbed inside the critical layer. Among convergent waves those only are absorbed which carries angular momentum of the same sign as one's of the rotation in the vortex. Convergent waves with the opposite sign of angular momentum are just reflected from the vortex. The absorption of a wave is possible only if the vortex flow is characterized by fast enough angular velocity there. The behavior of the wave near the critical layer can be described by the well-known model where the mean flow is rectilinear shear flow. We show that the conventional wave train approximation for the short-wave limit is not applicable in the vicinity of the layer and revise it, deriving the proper equation and reformulating the conservation law of wave action. For the vicinity of critical layer, a model which accounts for the viscous dissipation is derived; viscous effects are studied for absorption both of monochromatic wave and wave train.
\end{abstract}

\maketitle

\section{Introduction}

Dynamics of incompressible fluid in rotating systems is subjected to the Coriolis force additionally to inertia-viscous interplay~\cite{greenspan1968rotating}. Flow regime for which magnitude of the Coriolis force prevails over magnitude of the nonlinear interaction for the flow regime is quantified with low Rossby number $\mathrm{Ro}$. Rotating turbulence, which flows are characterized with high Reynolds $\mathrm{Re}$ and low Rossby numbers exhibits quasi-2D behavior. Its homogeneous form is the wave turbulence formed by an ensemble of interacting inertial waves. The ensemble has axisymmetric energy spectrum and gives preference at small scales to the waves having wave vectors almost perpendicular to the rotation axis~\cite{galtier2003weak,gelash2017complete,monsalve2020quantitative}. Stated quasi-2D behavior of the turbulence is manifested in the geostrophic flow component that is homogeneous along the rotation axis~\cite{davidson2013turbulence}. Dynamics of incompressible geostrophic flow is not affected by the Coriolis force  and it is stable against small vertically (i.e. along the rotation axis) inhomogeneous perturbations \cite{gallet2015exact}.

The geostrophic flow in form of long-living coherent vortices was observed in experimental setups with different kinds of small-scale forcing~\cite{godeferd2015structure}, e.g. forcing with multiple jets gushing from bottom~\cite{mcewan1976angular,ruppert2005extraction}, grid oscillating near the bottom~\cite{hopfinger1982turbulence} and forcing from the vertical side boundaries, homogeneous \cite{gallet2014scale} or inhomogeneous \cite{tumachev2023tworegimes,tumachev2024observation} along the vertical direction. The vertically inhomogeneous forcing produces predominantly inertial waves~\cite{greenspan1968rotating}, since the velocity field in the wave is vertically inhomogeneous as well, which spread over all the volume. Here we address mainly the last experimental realization from the list as it has implemented a number of different regimes of large-scale coherent columnar vortices.

The maintaining mechanisms for long-living geostrophic flow in turbulence still are not fully understood. In this paper we study in details the maintaining via inertial waves. Our goal is to sketch the mechanism of momentum and energy transfer from a wave to the geostrophic flow. The dispersion relation for inertial waves $\omega = 2s\Omega \cos\theta_{\textbf k}$, where $\Omega$ is angular velocity of the fluid rotation and $s$ is polarization of  wave~\cite{kraichnan1973helical}, has zeros, when the angle $\theta_{\textbf k}$ between the wave vector ${\textbf k}$ and the rotation axis $Oz$ becomes $\pi/2$. This fact combined with the Doppler effect imposed by inhomogeneous geostrophic flow with large enough velocity leads to existence of vertical critical layer, where the relative shifted frequency of the wave turns to zero and its wavenumber accordingly tends to infinity. The wave is absorbed by the mean flow near the layer, transferring to the latter its momentum and energy. Previously, the same absorption mechanism was explored in details for internal gravity waves in vertically stratified fluid with the dispersion relation $\omega = \pm N\sin\theta_{\textbf k}$, where $N$ is the Brunt-V\"{a}is\"{a}l\"{a} frequency~\cite{booker1967}. For internal waves critical layers are oriented horizontally. Recently, the theory was extended for the inertial waves in~\cite{baruteau2013inertial,astoul2021complex} considered in spherical geometry. 

The maintenance of large-scale coherent vortices by inertial waves was studied in \cite{kolokolov2020structure,parfenyev2021influence,parfenyev2021velocity} in assumption that they are excited on a background of unbounded uniform shear flow with some model forcing which is distributed homogeneously over all the volume. In this paper we consider cylindrical geometry and the forcing being localized in space, that allows one to systematically account for finite-size effects, when the wavelength and the geostrophic vortex size are comparable. In particular, this approach can be applied to the experimental realization \cite{tumachev2023tworegimes,tumachev2024observation}, when the vortex is maintained by convergent cylindrical waves coming from the peripheral region of the flow. We revisit the wave train theory~\cite{sutherland2010internal} and show that it gives wrong prediction for wave train width in a vicinity of the critical layer. Instead, we derive proper equation and in particular consider a wave train dynamics and show that the wave action~\cite{bretherton1968wavetrains,andrews1978wave} is still conserved here. Finally, we establish the influence of viscosity, which was previously considered in \cite{hazel1967heat&visc} for internal gravity waves.

The remainder of this paper is structured as follows. Wave propagation inside an axisymmetric steady vortex is considered in Section~\ref{sec:vort_flow}, where the wave equation is obtained and analyzed. In Section~\ref{sec:rect_flow} the model with rectilinear streamlines is considered in more details, which is suitable to describe wave behavior in the vicinity of the critical layer, including cylindrical geometry. The process of wave absorption in the vicinity of critical layer and wave reflection are considered in Section~\ref{sec:refl-or-abs} in approximation of the short-wave limit. The absorption process is considered more precisely in Section~\ref{sec:viscous_model}, where a proper consideration of wave train evolution and the viscosity influence are given. A discussion of the obtained results and a conclusion are given in Section~\ref{sec:conc}. Some technical details are moved aside into Appendices~\ref{app.sec:cylindrical_eq},\ref{app.sec:reflection},\ref{app.sec:viscous_eqs}.

\section{Wave Propagation inside Vortex} 
\label{sec:vort_flow}

A flow of incompressible fluid rotating along $Oz$ axis with the angular velocity $\bm\Omega=\Omega \bm e^z$ is governed by the Navier-Stokes equation with the Coriolis term:
\begin{equation}					\label{eq:NSE}
	\partial_{t}\bm{v}+ \left(\bm v,\nabla\right)\bm v+2\left[\bm{\Omega}\times\bm{v}\right]=-\nabla p+\nu\Delta\bm{v},\quad 
	\mathop{\mathrm{div}} \bm v=0.
\end{equation}
Here $\bm v$ is velocity field of the fluid, $p$ is the physical pressure per unit mass modified with centrifugal potential term, $\nu$ is kinematic viscosity coefficient and we choose $\Omega>0$ for definiteness. 

\begin{figure}[t]

\includegraphics[width=\linewidth]{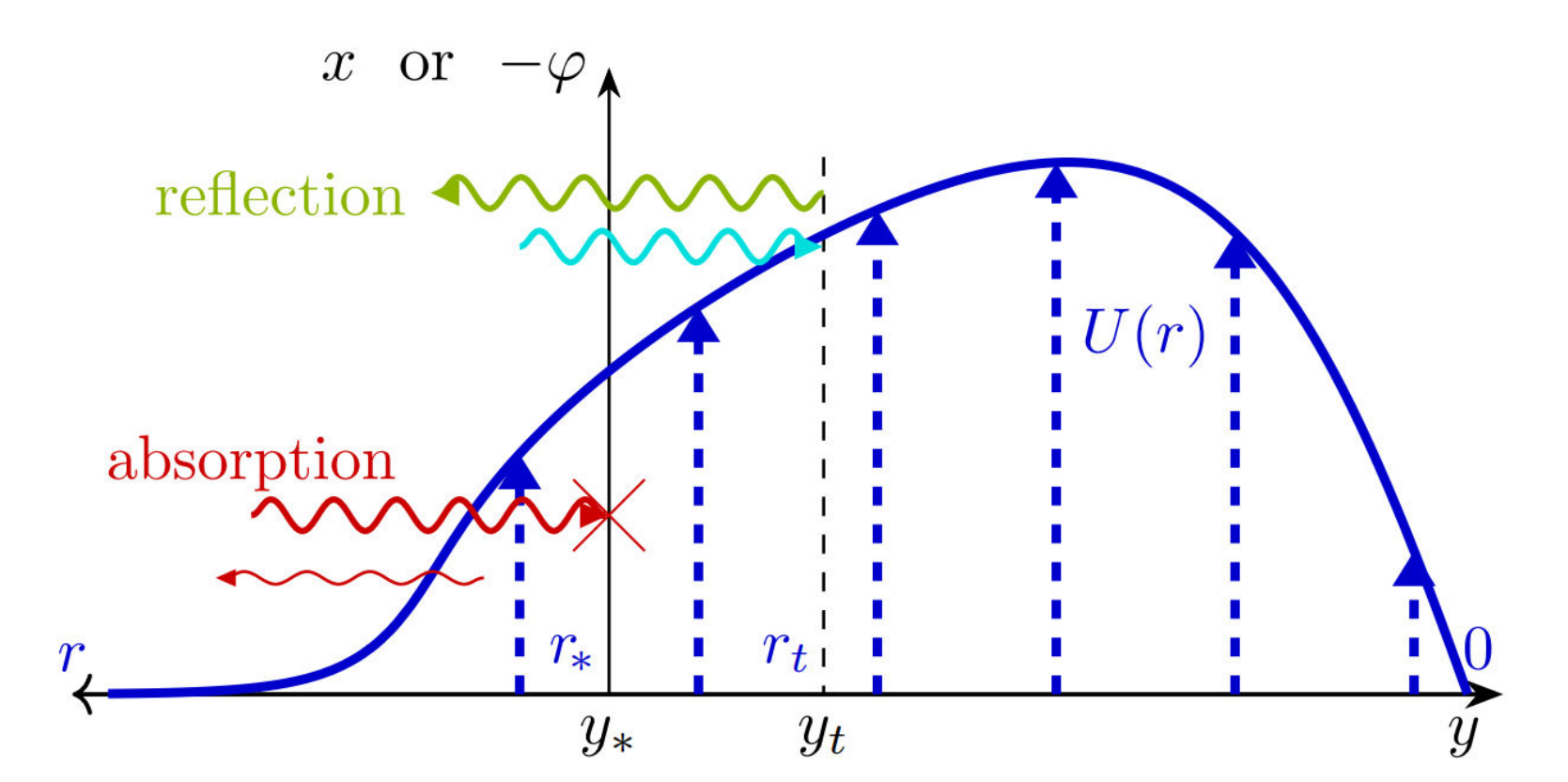} 

\caption{Schematics for inertial wave propagation: absorption with some reflection at the critical layer $r=r_\ast$ and total reflection at the reflection surface $r=r_t$. For the rectilinear model from Section~\ref{sec:rect_flow} we pass to the inverted $Oy$ axis, keeping subscript notations for special points.}

\label{fig:absorp_scheme}
\end{figure}

We consider propagation of inertial wave inside an axisymmetric geostrophic vortex. Thus, the velocity field can be decomposed in the vortex mean flow $\bm U$ that is constant in time, and the wave against its background, which is described by the velocity field $\bm u(t,\bm r)$, $\bm v=\bm U+\bm u$. In the cylindrical coordinate system $\left(r,\varphi,z\right)$, we consider the mean flow ${\bm U}$ as azimuthal and homogeneous in~$\varphi,z$:~$\bm U= U(r)\bm e^\varphi$. The vortex flow differs from rigid body rotation, if its angular velocity $U/r$ depends on radial position $r$. In this case, the flow is called differential rotation, which is characterized with nonzero local shear rate $\Sigma(r)=r\partial_r (U/r)$. We also presume that the vortex is localized in the transverse plane, $U(r)\rightarrow 0$ at $r\rightarrow\infty$, see Figure~\ref{fig:absorp_scheme} for illustrations. Wave velocity $\bm u =\left(u,v,w\right)$ vector components are: $\varphi$-component $u$ aligned with the mean flow, radial $r$-component $v$ that is the transverse one in the shear plane, and $z$-component $w$ aligned with the rotation axis. As the rotation is assumed to be fast, the Rossby number for motion of the wave is small, $|\nabla {\bm u}|/2\Omega\ll1$. In this limit, the rotation suppresses the nonlinear wave interaction which rate is of order of $|\nabla {\bm u}|^2/2\Omega$ \cite{galtier2003weak}. It is assumed to be smaller than $\Sigma$ as well, the latter is the nonlinear interaction rate between the wave and the geostrophic flow ${\bm U}$. So we use the linear in ${\bm u}$ approximation of (\ref{eq:NSE}) omitting the nonlinear term $\left(\bm u\cdot \nabla\right)\bm u$ to describe the wave motion:
\begin{align}\label{eq:lin-system_cyl}
	\nonumber
	\left(\partial_{t}+U/r\partial_{\varphi}\right) & v&-2\widetilde \Omega u &=-\partial_r p
    + 
    \nu(\Delta {\bm u})_r, 
    \\ \nonumber
	\left(\partial_{t}+U/r\partial_{\varphi}\right) & u  +{\Sigma v \hskip-15pt}&+2\widetilde \Omega v   &= -1/r\partial_\varphi p
    +
    \nu (\Delta {\bm u})_\varphi, 
    \\ 
	\left(\partial_{t}+U/r\partial_{\varphi}\right) & w & \quad 
    &= 
    -\partial_z p+ 
    \nu\Delta w.
\end{align}
Presence of local angular velocity of rotation of fluid element $\widetilde \Omega(r)=\Omega+U/r$ and of extended time derivative $\partial_{t}+U/r\partial_{\varphi}$ indicates the invariance of Eqs.~(\ref{eq:lin-system_cyl}) across the change in the global rotation frequency $\Omega$. It is assumed $\widetilde \Omega>0$ holds everywhere, i.e. the global rotation is not cancelled at some distance $r$ by the vortex rotation. The problem is homogeneous in $t,\varphi,z$ variables and allows one to study the solution in Fourier space, thus changing their derivatives with wave frequency $\omega$, which one may call `absolute' as is it measured in the reference system rotating with the fixed angular velocity $\Omega$, angle harmonic $m$ and wavenumber $k_z$ correspondingly:
\begin{equation}\label{homogeneity}
    \partial_{t}\rightarrow-i\omega,
    \quad
    \partial_{\varphi}\rightarrow i m,
    \quad
    \partial_{z}\rightarrow i k_z,
\end{equation}
where the axial number $m$ is integer.

In absence of the vortex flow, one derives from the system~(\ref{eq:lin-system_cyl}) in the inviscid limit the equation for pressure variance $p(\omega,r,m,k_z)$ of the inertial wave \cite{greenspan1968rotating}
\begin{equation}\label{eq:in-waves_bare}
	\omega^{2}\left(\partial^2_{r}+ \frac{1}{r}\partial_{r}-\frac{m^{2}}{r^{2}}-k_z^{2}\right)p+4\Omega^{2}k_z^{2}p=0,
\end{equation}
which supports two polarizations for each wavevector in the cylindrical coordinates:
\begin{equation}\label{dispersion01}
    p(t,\bm{r})=J_{m}\left(qr\right)e^{im\varphi+ik_zz-i\omega_{s}t},
    \quad
    \omega_s=\frac{2s\Omega k_z}{k},
\end{equation} 
where the wavenumber $k = \sqrt{ k_z^{2}+q^{2}}$ is positive by definition and $s=\pm1$ is the sign of polarization.

In the presence of the vortex flow, still neglecting the viscosity, we derive equations that modify this free-wave model, the procedure is carried out in Appendix \ref{app.sec:cylindrical_eq}. It is natural to introduce the relative wave frequency $\widetilde{\omega}(r)$ in the reference frame moving with an element of fluid
\begin{equation}\label{eq:Doppler_freq}
	\widetilde{\omega}(r)=\omega-mU(r)/r,
\end{equation}
that accounts the Doppler shift at radius $r$~\cite{sutherland2010internal}. The relations between $\bm u$ components and $p$ in terms of $\left(\omega, r, m, k_z\right)$ are
\begin{gather}
                                    \label{cyl:u-via-v}
    u=\frac{im/r}{k_{\scriptscriptstyle\perp}^{2}}\left(v^{\prime}+v/r\right)-\frac{ik_z^{2}}{k_{\scriptscriptstyle\perp}^{2}}\frac{2\widetilde{\Omega}+\Sigma}{\widetilde{\omega}}v,
    \\                              \label{cyl:w-eq}
    w=\frac{ik_z}{k_{\scriptscriptstyle\perp}^{2}}\left(v^{\prime}+v/r\right)+\frac{ik_zm/r}{k_{\scriptscriptstyle\perp}^{2}}\frac{2\widetilde{\Omega}+\Sigma}{\widetilde{\omega}}v,
    \\
    \widetilde{\omega}w=k_z p,      \label{cyl:p-eq}
\end{gather}
where  $k_{\scriptscriptstyle\perp}^2(r)=k_z^{2}+m^{2}/r^{2}$ and prime stands for derivative with respect to $r$. The wave equation for $p$ becomes rather complicated compared to (\ref{eq:in-waves_bare}), see Eq.~(\ref{eq:p-eq_cyl}) in Appendix \ref{app.sec:cylindrical_eq}. Further in paper we will work with the equation for radial component $v$, which is
\begin{equation}\label{eq:v-eq_cyl}
    \begin{aligned}
    \widetilde{\omega}^{2}\left(\frac{d}{dr}\left(\frac{v^{\prime}+v/r}{k_{\scriptscriptstyle\perp}^{2}}\right)
    -
    v\right)
    + 
    \frac{2\widetilde{\Omega}\left(2\widetilde{\Omega}+\Sigma\right)k_z^{2}}{k_{\scriptscriptstyle\perp}^{2}}v
    \ +
    &
    \\[10pt] 
    +\ 
    \widetilde{\omega} v\left(\frac{d}{dr}-\frac{1}{r}\right)\frac{m/r\left(2\widetilde{\Omega}+\Sigma\right)}{k_{\scriptscriptstyle\perp}^{2}}
    =
    0.&
    \end{aligned}
\end{equation}
For description of non-monochromatic wave field dynamics one should substitute $\widetilde \omega \to i\partial_t - m U/r$ in it. 

Consider a monochromatic wave which is a solution of (\ref{eq:v-eq_cyl}). Such wave does not exchange angular momentum and energy with the mean flow, till the wave encounters a critical layer, which will be discussed in details in Subsection~\ref{subsec:invisc_abs}. Indeed, the flux density of $\varphi$-component of momentum in $r$-direction associated with the wave motion is
\begin{equation}\label{eq:momentum_flux}
    \Pi_{r\varphi}
    =
    \big\langle v_r v_\varphi\big\rangle_{\!\mathrm{w}}
    =
    \frac{\mathrm{Re}\big(uv^{\ast}\big)}{2}
    =
    -\frac{m\mathrm{Im}\big(v^{\prime}v^{\ast}\big)}{2rk_{\scriptscriptstyle\perp}^{2}}
    ,
\end{equation}
where angle brackets $\langle \ldots \rangle_{\!\mathrm{w}}$ denote averaging in time over the wave oscillations and the relation (\ref{cyl:u-via-v}) was used. The~mean energy flux  density in $r$-direction associated with the wave is proportional to the flux of momentum,
\begin{equation}\label{eq:energy_flux}
    j_{r}^{\scriptscriptstyle E}
    =
    \left\langle \left(\frac{{\bm v}^2}{2}+p\right)v_r\right\rangle_{\!\!\mathrm{w}}
    =
    \frac{\mathrm{Re}\big(p v^{\ast}+Uuv^{\ast}\big)}{2}
    =
    \frac{\omega r}{m}\Pi_{r\varphi},
\end{equation}
according to Eq.~(\ref{cyl:u-via-v},\ref{cyl:w-eq},\ref{cyl:p-eq}).  The total flux of wave's angular momentum (its vertical component) is $r^2 \Pi_{r\varphi}$, and the total energy flux in radial direction is $rj_{r}^{\scriptscriptstyle E}$ in cylindrical coordinates. They do not depend on $r$,
\begin{equation} 				\label{eq:cyl_fluxes}
    \left(r^2 \Pi_{r\varphi}(r)\right)^\prime=0,
    \qquad 
    \left(  r j^{\scriptscriptstyle E}_r(r)\right)^\prime =0,
\end{equation}
according to Eq.~(\ref{eq:v-eq_cyl}), since both $v$ and $v^{\ast}$ solve it. Thus, the volume force applied by the wave to the mean flow in the azimuthal $\varphi$-direction is zero, see e.g.~\cite{kolokolov2020structure,parfenyev2021influence}.

Next, consider the short-wave limit, when the wavelength is small compared to the characteristic scale of the vortex flow $U(r)$. Keeping main terms in (\ref{eq:v-eq_cyl}), we arrive to 
\begin{equation}\label{eq:lin-system_cyl1}
    \widetilde\omega^2 \big(v^{\prime\prime} - k_{\scriptscriptstyle\perp}^{2} v\big) 
    +
    2\widetilde{\Omega}\big(2\widetilde{\Omega}+\Sigma\big)k_z^{2} v
    =
    0.
\end{equation}
If $v$ corresponds to a travelling wave which moves inward or outward from the vortex's center, then $v^{\prime\prime} \approx - k_r^2 v$, where $k_r$ is the radial wavenumber. Equation (\ref{eq:lin-system_cyl1}) yields the dispersion relation 
\begin{equation}\label{dispersion02}
    \widetilde \omega
    =
    \frac{s\sqrt{2\widetilde{\Omega}(2\widetilde{\Omega}+\Sigma)}\,k_z}{k},
    \qquad 
    k^2 = k_r^2 + k_{\scriptscriptstyle\perp}^{2},
\end{equation}
which is a generalization of (\ref{dispersion01}). It should be $2\widetilde\Omega+\Sigma > 0$ in order to the wavenumber $k$ was real. The propagation of an inertial wave in a shear flow that is homogeneous and unbounded in space was considered in \cite{salhi1997analysis}, and the limit of weak shear rate $\Sigma/2\widetilde\Omega\ll1$ was considered in \cite{kolokolov2020structure}. These studies regarded an inertial wave as plane one in unbounded space, so its wavevector evolves in time on a characteristic there. In contrary, here we consider homogeneous evolution in time, but inhomogeneous wave field distribution in space. A combination of (\ref{eq:Doppler_freq}) with (\ref{dispersion02}) gives
\begin{equation}\label{dispersion03}
    \frac{\sqrt{1+\Sigma/2\widetilde \Omega}\ \widetilde\Omega/\Omega}{k}
    =
    \frac{1}{k_\infty} 
    - 
    \frac{sm \cdot\big(U/r\big)}{2\Omega k_z}.
\end{equation}
Here $k_\infty$ is the wavenumber at infinity, where $U/r=0$ and $\Sigma=0$, so $\omega =2 s\Omega k_z/k_\infty$. 

If $\omega mU >0$, which is for a wave with polarization $smU/k_z>0$ in (\ref{dispersion03}), and the magnitude of the vortex flow $U$ is large enough, then there is a cylindrical surface~$r=r_\ast$, where $k\to\infty$, since $\widetilde{\omega}=0$ according to~(\ref{dispersion03}). Such surface is called the critical layer after~\cite{booker1967}. Equation (\ref{eq:Doppler_freq}) means a match at $r=r_\ast$ between the frequency of an excited inertial wave and the Doppler shift it acquired due to propagation in the mean flow. The surface is singular for the inviscid wave equation (\ref{eq:v-eq_cyl}) as well. Before the layer, the radial wavenumber $k_r$ tends to infinity together with $k$, so the wavelength in the radial direction is shortened. In order to study the wave behavior near its critical layer it is sufficient to pass to the simplified model with local Cartesian coordinate system, see in Figure~\ref{fig:absorp_scheme}, thus dropping out of our consideration that streamlines of the mean shear flow are curvilinear. From the study presented in Subsection~\ref{subsec:invisc_abs} it follows that the wave amplitude attenuates exponentially after the passing of the critical layer. Therefore, equation (\ref{eq:cyl_fluxes}) is not valid at the critical layer, and the wave transfers its angular momentum and energy to the mean flow.

For the other polarization such that $\omega mU <0$ or, equivalently, $smU/k_z<0$ the wavenumber $k(r)$ diminishes with increase of vortex's flow in magnitude. There is a cylindrical surface $r=r_t$, where the wavenumber reaches its minimum possible value, $k=k_{\scriptscriptstyle\perp}$, so the radial wavenumber is zero, $k_r=0$. The surface $r=r_t$ is not a singularity in the wave equation (\ref{eq:v-eq_cyl}), it corresponds to a turning point in WKB approximation, see Subsection~\ref{subsec:short-wave-limit}. A convergent cylindrical wave is reflected from the surface, thus turning to divergent one. There is no any angular momentum transfer in the case. Figure~\ref{fig:absorp_scheme} illustrates the absorption and reflection of waves. The reflection surface $r_t$ is present for waves with the opposite polarization $smU/k_z>0$ as well. If the vortex flow is weak enough, then a convergent wave with $smU/k_z>0$ meets its reflection surface first, not a critical layer, because its position is closer to the vortex axis, $r_\ast < r_t$. There is no angular momentum transfer from such wave too. (The wave's amplitude decays with the distance after it passes the reflection surface, so it becomes attenuated significantly before the critical layer. Thus, the transferred angular momentum transfer is negligible in this case.)

Finally, the Doppler effect does not affect the wave with axial symmetry ($m=0$). For this wave's dynamics mean flow only modifies the angular frequency as $2\widetilde\Omega (2\widetilde\Omega + \Sigma)$ in (\ref{eq:v-eq_cyl},\ref{eq:lin-system_cyl1}). Convergent wave passes the entire vortex right up to its axis and then it is reflected backwards. We note that stability of Couette flow against axisymmetric perturbations was studied in \cite[\S~15]{drazin_reid_book2004}.

\section{Model with rectilinear streamlines} 				\label{sec:rect_flow}

As it is discussed above, one can pass to local Cartesian coordinate system for description of inertial wave behavior near its critical layer or reflection layer in the short wave limit. Here we reformulate the problem of inertial wave propagation in a shear flow with rectilinear streamlines first. Henceforth, we consider that an inertial wave with velocity field $\bm u$ enters into the domain of constant flow $\bm U= U(y){\bm e}^x$ from outside: asymptotically at $y\rightarrow -\infty$ mean flow is absent, $U\rightarrow 0$, see Figure~\ref{fig:absorp_scheme}. The linearized Navier-Stokes equation for the wave dynamics against the nonzero $U(y)$ corresponds to the system~(\ref{eq:lin-system_cyl}) in zero curvature limit:
\begin{equation}					\label{eq:linear_wave}
    \big(\partial_{t}+U\partial_{x}\big)\bm{u}
    +
    u_{y}{\bm e}^{x}U^\prime
    +
    2\left[\bm{\Omega}\times\bm{u}\right]
    =
    -\nabla p + \nu \Delta{\bm u}.
\end{equation}
We adapt the components notation from the cylindrical geometry: $u_x\equiv u$ for component in the shear direction, $u_y\equiv v$, $u_z\equiv w$.

First, let us neglect viscous effects and consider the wave in the region $y\to-\infty$, where $\bm u$ is implied to be excited as pure inertial wave, which is a solution of the equation
\begin{equation}
    \partial_t^2 \Delta \bm u
    + 
    4\left(\bm \Omega\cdot \nabla\right)^2\bm u
    =0,
    \qquad 
    \mathop{\mathrm{div}} {\bm u}=0.
\end{equation}
The velocity vector in a monochromatic plane wave with wave vector ${\bm k}$ lies in the plane transverse to ${\bm k}$, which we provide with Craya–Herring basis~\cite{herring1974,sagaut2008homogeneous}:
\begin{equation}
    {\bm e}^1 
    = 
    \frac{\left[\bm k \times {\bm e}^z \right]}{\left| \left[\bm k \times {\bm e}^z \right]\right|},
    \qquad 
    {\bm e}^2 = \frac{\left[\bm k\times {\bm e}^1\right]}{k}.
\end{equation}
It is convenient to use the velocity vector expansion in the basis $\bm h_{\bm k,s}$ of circular (or helical) polarizations~\cite{kraichnan1973helical}:
\begin{equation}					\label{eq:wave_asympt}
    \left.\bm u\right|_{y\rightarrow-\infty}
    =
    b_{\bm k,s} \bm h_{\bm k,s} 
    e^{-i\omega_s t+i \bm k \cdot \bm r},
    \qquad 
    \bm{h}_{\bm{k},s}=\frac{{\bm e}^2-is {\bm e}^1}{\sqrt{2}},
\end{equation}
where the frequency $\omega_s$ is that defined in (\ref{dispersion01}).

For a wave in a nonzero shear flow, we keep the notations for the vertical wavenumber $k_z$ and the global frequency $\omega$ from~(\ref{homogeneity}), but change the azimuthal wavenumber to $\partial_x \to ik_x$. One can reduce system (\ref{eq:linear_wave}), excluding $w, p$,  and continue with the components on the shear plane, using the incompressibility condition and the third component of Euler equation:
\begin{align} 					\label{eq:uz}
    	k_z w & =  -k_x u+i v^\prime  
    \\							\label{eq:p_nonvisc}
    	\left(\omega-k_x U\right)w
    &=
    k_z p,
\end{align}
where prime denotes the derivative in $y$. Substituting it in $x$-component of Euler equation (\ref{eq:linear_wave}), we express $u$ as:
\begin{equation}\label{eq:ux}
    u
    =
    \frac{ik_{x}v^{\prime}(y)}{k_{\scriptscriptstyle\perp}^{2}}
    +
    \frac{i\left(2\Omega+\Sigma\right)k_{z}^{2}v}
    {k_{\scriptscriptstyle\perp}^{2}\widetilde{\omega}},
\end{equation}
where $\widetilde{\omega}(y) = \omega -k_x U(y)$, $k_{\scriptscriptstyle\perp}=\sqrt{k_x^2+k_z^2}$ are redefined for the rectilinear model and now $\Sigma = -U^{\prime}$ due to the change of coordinates' order in $(x,y,z)$ frame. Note that if the large-scale Rossby number is small, $\Sigma/2\Omega\ll1$, then  the relation between the velocity components given by (\ref{eq:uz},\ref{eq:p_nonvisc},\ref{eq:ux}) corresponds to those set by (\ref{eq:wave_asympt}) for plane wave.  The last $y$-component of (\ref{eq:linear_wave}) becomes a wave equation \cite{astoul2021complex}:
\begin{equation}					\label{eq:v_invisc}
    \widetilde{\omega}^{2}\big(v^{\prime\prime}
        -k_{\scriptscriptstyle\perp}^{2} v\big)
    +
    \Big(2\Omega\big(2\Omega+\Sigma\big)k_{z}^{2}
    -
    \widetilde{\omega}k_{x}\Sigma^{\prime}\Big)v
    =
    0,
\end{equation}
which can be obtained form (\ref{eq:v-eq_cyl}) in the limit $r\to \infty$, but with the fixed wavenumber in the streamwise direction $m/r = k_x$. The Eq.~(\ref{eq:v_invisc}) is similar to the Taylor-Goldstein equation describing internal gravity waves in stratified fluid~\cite{miles1961, howard1961, booker1967}, which is derived from the Boussinesq system.

For a monochromatic wave, components of its momentum flux $\Pi_{xy}$ and energy flux $j^{\scriptscriptstyle E}_y$ are determined by the same equations (\ref{eq:momentum_flux},\ref{eq:energy_flux}) with changes $v_\varphi\to v_x$, $v_r\to v_y$ and $m/r\to k_x$: 
\begin{equation}\label{Pi-jE-rectilinear}
    \Pi_{xy} = 
    -\frac{k_x\mathrm{Im}\big(v^{\prime}v^{\ast}\big)}{2k_{\scriptscriptstyle\perp}^{2}},
    \qquad 
    j^{\scriptscriptstyle E}_y
    =
    \frac{\omega}{k_x} \Pi_{xy},
\end{equation}
that can be checked directly from (\ref{eq:uz},\ref{eq:p_nonvisc},\ref{eq:ux}). In the rectilinear geometry, the fluxes are constant, 
\begin{equation}
    \partial_y \Pi_{xy}=0,
    \qquad 
    \partial_y j^{\scriptscriptstyle E}_y =0.
\end{equation} 
This statement can be checked using the wave equation~(\ref{eq:v_invisc}).

\subsection{Short-wave limit}
\label{subsec:short-wave-limit}

Now we consider the short-wave limit, in which the wavelength is much smaller than scales of change in the mean flow  $U(y)$. The solution for a travelling monochromatic wave within WKB approximation is 
\begin{align}\label{quasiclassic-sol}
    v&(t,y) 
    =
    \Phi
    \exp\left(-i\omega t \pm i \int\limits^y d \xi \, k_y(\xi) \right),
\end{align}
where  $k_y(y)$ satisfies the dispersion relation that matches with Eq.~(\ref{dispersion02}) in cylindrical coordinates after change in coordinate from $r$ to $y$: $k_r\to k_y$, $\widetilde \Omega (r)\to \Omega$ and $m/r\to k_x$, so $k_{\scriptscriptstyle\perp}$ becomes constant. Here $\Phi(y) = a/\sqrt{k_y(y)}$ is the complex envelope, and the constant $a$ stands for complex wave amplitude. The approximation holds valid if the condition
\begin{equation}					\label{quasiclassic-cond}
	k_y^\prime(y)\ll \left(k_y(y)\right)^2
\end{equation}
is satisfied throughout its movement. The momentum flux (\ref{Pi-jE-rectilinear}) is now 
\begin{equation}    				\label{eq:momentum_flux1}
    \Pi_{xy}
    =
    -\frac{k_xk_y |a^2|}{2k_{\scriptscriptstyle\perp}^{2}|k_y|}
    ,
\end{equation}

Although a monochromatic wave does not exchange momentum and energy with the mean current, this does not mean that a non-stationary wave train also travels without the exchange. The theory of a wave train propagation in context of inertial gravity waves is presented e.g. in book~\cite{sutherland2010internal}. To reproduce the procedure for inertial waves, we consider $\Phi$ to be a function of both time and coordinate, which is slowly varying compared to the exponent in (\ref{quasiclassic-sol}) (choosing one with a plus sign). Also we restore the time derivative in the wave equation (\ref{eq:v_invisc}), $\widetilde \omega \to i\partial_t - k_x U$. To obtain the equation governing the dynamics of envelope $\Phi$ in the main approximation, one should account for the first derivatives of slowly varying $k_y$, $\Sigma$ and $\Phi$:
\begin{equation}\label{envelope}
    \left(\partial_t + {\mathrm v}_{\!g\,} \partial_y + \frac{{\mathrm v}_{\!g\,}}{2k_y} k_y^\prime\right)\Phi
    =
    \frac{ik_x \Sigma^\prime}{2k^2}\Phi,
\end{equation}
where $k(y)=\sqrt{k_y^2+k_{\scriptscriptstyle\perp}^2}$ and the group velocity in $y$-direction is
\begin{equation}\label{group-velocity-rectilinear}
    {\mathrm v}_{\!g\,}
    =
    \frac{\partial\widetilde\omega_s}{\partial k_y}
    =
    -\frac{k_y \widetilde\omega}{k^2}.
\end{equation}
As it should be, the sign of the energy flux $j^{\scriptscriptstyle E}_y$, see~(\ref{Pi-jE-rectilinear}), coincides with the sign of the group velocity.

The right-hand side of the equation~(\ref{envelope}) arises from relatively small last term in~(\ref{eq:v_invisc}) and only yields extra phase change for the envelope, so it is not of our interest now. Concerning the absolute value of the envelope, equation  (\ref{envelope}) can be rewritten in the form 
\begin{equation}\label{divergence-vg}
    \big(\partial_t + \partial_y {\mathrm v}_{\!g\,}\big)
    \frac{k_y|v|^2}{{\mathrm v}_{\!g\,}}
    =0. 
\end{equation}
Using the relation $k_{\scriptscriptstyle\perp}^2 |{\bm u}^2| = (2+\Sigma/2\Omega)k^2 |v|^2$ which is obtained from (\ref{eq:uz},\ref{eq:p_nonvisc},\ref{eq:ux}), where differentiation with respect to $y$ accounts for the fast oscillating exponent in (\ref{quasiclassic-sol}) only, we arrive to the conservation of the integral quantity
\begin{equation}\label{wave-action}
    \int \frac{|{\bm u}^2|\,dy}{(1+\Sigma/4\Omega)\, \widetilde \omega}
    =
    \mathrm{const}.
\end{equation}
The formula~(\ref{wave-action}) stands for the conservation of wave action, that is a general law for waves travelling against the background of moving media, see \cite{sutherland2010internal, bretherton1968wavetrains,andrews1978wave}. As $\widetilde \omega$ is not constant in a vortex flow, the conservation law (\ref{wave-action}) means that a wave train's energy in not conserved in time. The total energy of full flow ${\bm v}$ is conserved because we consider Euler equation. Thus, the wave train exchanges energy with the mean flow. 

\section{Wave reflection and absorption} 						\label{sec:refl-or-abs}

To consider only those waves that move from the vortex peripheral region toward its axis, we should assume that the energy flux (\ref{Pi-jE-rectilinear}) is positive. As sign of the relative wave frequency $\widetilde \omega$ coincides with the sign of the global wave frequency $\omega$, we arrive to the conditions: 
\begin{equation}\label{vg-pis-cond}
    j^{\scriptscriptstyle E}_y > 0,
    \qquad 
    \mathop{\mathrm{sign}} \Pi_{xy}
    =
    \mathop{\mathrm{sign}} \big(\omega k_x\big).
\end{equation}
For simplicity, we assume $U(y)$ monotonically increases its absolute value along $Oy$ axis, so the product $\Sigma U < 0$.

\subsection{Wave reflection and trapping}
\label{subsec:reflection}

Consider those waves which relative frequency $\widetilde\omega$ increases in magnitude when approaching the vortex center, that~is 
\begin{equation}\label{reflection-cond}
    \omega k_x U < 0. 
\end{equation}
If the mean flow is strong enough, there is a surface $y=y_t$, where $k_y=0$ according to the dispersion relation with the Doppler shift~(\ref{dispersion02}). Let $\eta = y_t - y$ and $\widetilde \omega(y_t) = \widetilde\omega_t$, then the monochromatic wave equation near the surface satisfies the Airy equation 
\begin{equation}\label{Airy}
    \eta_t^3 v^{\prime\prime}
    +
     \eta v
    =
    0,
    \qquad 
    \eta_t^3 = 
    \left(\frac{\widetilde  \omega_t}{2\Sigma k_x}\right)
    \frac{1}{k_{\scriptscriptstyle\perp}^2},
\end{equation}
and the factor in the round brackets is positive according to (\ref{reflection-cond}). Eq.~(\ref{Airy}) describes reflection of a wave which comes from large negative $y$ and reflects near the point $y=y_t$ back to the region \cite[\S\,4.2]{berry1972WKB}. The region $y>y_t$ is forbidden for classical motion in WKB approximation, so the wave decays there. 

The solution $v\propto \mathrm{Ai}(-\eta/\eta_t)$ of (\ref{Airy}) that decays at $y>y_t$, can be chosen real and treated as standing wave, consisting of incident and reflected travelling waves with equal amplitudes. Thus, the reflection is elastic and total momentum flux $\Pi_{xy}$ (\ref{eq:momentum_flux}) associated with the wave is zero. For the incident wave it has a sign opposite to $U$, $U\Pi_{xy}<0$. In cylindrical geometry this means such waves transfer the angular momentum opposite to the angular momentum of the vortex flow. Note also that the reflection conserves the frequency $\omega$ and wave wavenumber $k_z$, so the sign $s$ of polarization remains the same. 

Reflection of a wave train lasts a finite time. Indeed, consider the region $\eta\gg |\eta_t|$ where the short-wave limit is achieved. The wavenumber $k_y=\pm \sqrt{\eta/\eta_t^3}$ and the group velocity (\ref{group-velocity-rectilinear}) ${\mathrm v}_{\!g} \sim \eta^{1/2}$ in the region, so the traveling time to/from the reflection point $\int d \eta/{\mathrm v}_{\!g}$ converges at small distances. In the vicinity of the reflection point, at $\eta\sim\eta_t $, the kinetic energy density increases as ${\bm u}^2 \propto \eta^{-1/2}$ up to some proximity of zero. Due to the reflection is elastic, the wave action (\ref{wave-action}) integral for the reflected wave train equals to the one that the incident train had previously at the same location. Thus, the reflection process leads to reversible in time energy exchange between the waves and the mean flow.

For the internal gravity waves problem, a more complicated situation is also notable, when a wave is present in a classically allowed region that is restricted by turning points $y_{t1}$ and $y_{t2}$ from both sides. Such possibility of wave trapping in the shear flow was examined in~\cite{badulin1985trapping,badulin1993irreversibility}, being thought there it is caused by vertical inhomogeneity of the Brunt-V\"{a}is\"{a}l\"{a} frequency $N(z)$ in stratified fluid. For the problem of inertial waves in cylindrical geometry, local angular velocity $\widetilde\Omega(r)$ may play this role, see~(\ref{dispersion02}).

\subsection{Absorption at critical layer in inviscid limit}
\label{subsec:invisc_abs}

In the opposite case, 
\begin{equation}\label{omega-condition-critical}
    \omega k_x U > 0,
\end{equation}
the wave's relative frequency $\widetilde\omega(y)$ decreases and turns to zero at some layer $y=y_\ast$, $\widetilde\omega(y_\ast)=0$, if the magnitude of the mean flow $U$ is large enough. For this situation, it will be shown below that constant values of fluxes~(\ref{Pi-jE-rectilinear}) differ on each side from the critical layer. The difference means an exchange of energy and momentum between waves and the mean flow. Note that whereas the relative frequency $\widetilde \omega$ changes its sing when passing the critical layer, $k_z$-wavenumber remains unchanged. Hence, the dispersion law (\ref{dispersion02}) leads to inversion of the wave polarization $s$ when passing the critical layer.

In the vicinity of layer we introduce shifted coordinate $\eta=y-y_\ast$, updating the notation, and approximate mean velocity field with linear expansion, so $\widetilde{\omega}\approx \Sigma k_{x}\eta$, where $\Sigma\equiv\Sigma(y_\ast)$ is taken at the critical layer. There monochromatic wave is a solution of approximated Eq.~(\ref{eq:v_invisc}) that is characterized with the local Rossby number $\rho$:
\begin{equation}					\label{eq:v-rho}
    v^{\prime\prime}+\frac{1}{\rho^2\eta^2}v=0, 		
    \quad 
    \rho
    =
    \frac{\big|\Sigma k_{x}/k_{z}\big|}
    {\sqrt{2\Omega (2\Omega+\Sigma)}}.
\end{equation}
The equation (\ref{eq:v-rho}) is valid around the layer for $\eta$ below the characteristic scale $\eta_c =\left|\Sigma/ U^{\prime\prime}\right|\cdot \mathrm{min}\,(1,\rho^{-2})$, where the curvature of the velocity profile $U^{\prime\prime}\equiv U^{\prime\prime}(y_\ast)$ is also taken at the critical layer. Solutions for $v(\eta)$ have power-law dependency at the origin:
\begin{equation}\label{eq:zero_asympt}
    v_\pm\sim\eta^{(1\pm i \beta)/2},
    \quad 
    \beta
    =
    \left.\sqrt{4-\rho^2}\middle/\rho\right.,
\end{equation}
that performs neutral oscillating behavior in case $\rho<2$.

Let's draw a parallel to the problem of internal gravity waves in stratified fluid as perturbations to the shear flow $U(z)\bm{e}_x$ that is uniform in horizontal plane. One writes analogues to~(\ref{eq:v-rho},\ref{eq:zero_asympt}) for vertical velocity component $w(z)$ with the expression for parameter $\beta_N$ in exponent: $\beta_N^2= 4(Nk_{\scriptscriptstyle\perp}/\Sigma k_x)^2-1$, where $N$ is the Brunt-Väisälä frequency. It has a lower bound $\mathrm{inf}\beta_N=\sqrt{4\mathrm{Ri}-1}$ via the local Richardson number $\mathrm{Ri}=N^2/\left(U^\prime\right)^2$ that is independent of wavenumbers.  The inequality $\mathrm{Ri}(z_\ast)>1/4$ ensures neutral behavior at $z=z_\ast$ plane regarding to any perturbations of the shear flow in linear analysis. The sufficient condition for stability of the parallel shear flow in stratified fluid is that the Richardson number holds greater than~$1/4$ throughout the flow. This stability condition is known as the Miles–Howard theorem~\cite{miles1961,howard1961}.  The analysis for the similar problem of the critical layer for internal waves is presented in~\cite{booker1967}. One can find examples of this problem for particular models of the mean flow profiles in Refs.~\cite{miles1967,drazin1979,duin1982}.

Now, we take $v (\eta)$ in (\ref{eq:zero_asympt}) as the solution describing inertial wave that passes through the critical layer to the right, in the region $\eta>0$. For the solution $v_{\pm}$ of (\ref{eq:zero_asympt}) the local wavenumber is $k_y=\pm \beta /(2\eta)$; together with the relative frequency $\widetilde\omega=\Sigma k_x \eta$ it gives the sign of group velocity $\mathrm{sign}\,\mathrm{{\mathrm v}_{\!g}}=\mp\mathrm{sign}\,\Sigma k_x$ respectively, see~(\ref{group-velocity-rectilinear}). Thus, we choose the solution
\begin{equation}                \label{eq:02}
    v_\sigma \sim \eta^{(1+i\sigma \beta)/2},
    \qquad 
    \sigma = -\mathrm{sign}\,\Sigma k_x .
\end{equation}
For monotonous flow profile $U(y)$ the parameter $\sigma$ stays for the sign of frequency $\omega$, see comment below~(\ref{vg-pis-cond}).

To continue solution from the region $\eta>0$ to the region $\eta<0$, we use causality in time concerns that imply the regularization of (\ref{eq:v_invisc}) singular point. The regularization in complex plane of Fourier variable $\omega$ demands that the singularity shifts toward the half plane $\mathop{\mathrm{Im}}\omega<0$. It sets the branch of the power function in (\ref{eq:zero_asympt}) by the rule:
\begin{equation}				\label{eq:crit_regularization}
    \eta 
    \rightarrow 
    \eta -0i\sigma.
\end{equation}
Further in Subsection~\ref{subsec:visc_wave} we will arrive to the same regularization by accounting viscous dissipation in the model. For the solution (\ref{eq:02}) one obtains the relation between its values on different sides from the critical layer,
\begin{equation}                \label{eq:phase_change}
    \left.\eta^{(1+i\sigma\beta)/2}\right|_{\eta<0}=-i\sigma e^{\pi\beta/2}\left.\eta^{(1+i\sigma\beta)/2}\right|_{\eta>0}.
\end{equation}
The result can be expressed in amplitude of a transmitted through the critical layer wave:
\begin{equation}				\label{eq:transmission}
    t= i \sigma e^{-\pi \beta/2}.
\end{equation}
We note that the amplitude of the other wave $v_{-\sigma}$ is, conversely, larger from the side of positive $\eta$, though its group velocity is directed to the left. Therefore, a wave passing the critical layer is attenuated in the region it is moving to. The attenuation of the momentum and energy fluxes (\ref{Pi-jE-rectilinear}) for transmitted wave is
\begin{equation}	\label{eq:crit_absorp}
	\frac{\mathrm{Im}\left.v_{\sigma}^{\prime}v_{\sigma}^{\ast}\right|_{\eta>0}}{\mathrm{Im}\left.v_{\sigma}^{\prime}v_{\sigma}^{\ast}\right|_{\eta<0}}=e^{-\pi\beta}.
\end{equation}

Consider the limit of small $\rho$, that is $\beta\approx 2/\rho\gg1$. The limit $\rho\ll 1$ is a local manifestation of applicability condition~(\ref{quasiclassic-cond}) for WKB approximation at the critical layer. For the majority of waves in ensemble that enters the vortex flow it is reasoned with the limit of large $\Omega$. The flux ratio (\ref{eq:crit_absorp}) is exponentially small in the limit. Being discontinuous at $\eta=0$, the attenuation of $\Pi_{xy}$ means the exchange between wave and the shear flow at the vicinity of critical layer. In $\rho\ll 1$ limit the wave produces shear force $-\partial_y \Pi_{xy}=- \Pi_{xy}^{(0)}\delta(y-y_\ast)$, transferring effectively almost all its momentum as well as energy to the mean flow~\cite{tumachev2024observation}. 

The solution (\ref{eq:zero_asympt}) can be continued as a one-way travelling wave on $Oy$, if the WKB approximation~(\ref{quasiclassic-sol}) is valid all over the vortex. However, partial reflection of the incident wave during its propagation in the shear flow~$U(y)$ stays beyond the WKB approximation, being supressed if~(\ref{quasiclassic-cond}) holds. The reflection reduces efficiency of the momentum and energy transfer from the incident wave to the mean flow. It is shown in Appendix~\ref{app.sec:reflection} that a problem of wave propagation to its critical layer point $y_\ast$ can be reformulated as the scattering theory for waves in the form of WKB-expression~(\ref{quasiclassic-sol}). In the limit (\ref{quasiclassic-cond}), the relative amplitude $r$ of the reflected wave in the Born approximation~(\ref{refl:Born-r_coeff}) is 
\begin{equation}                \label{eq:reflection}
    r
    =
    -\frac{1}{4}\intop_{-\infty}^{y_{\ast}}dy\exp\left(2i\sigma\intop_{y_{0}}^{y}d\xi\, k_{y}(\xi)\right)\mathcal{V}^{\prime}(y),
\end{equation}
where $k_y(y)$ is the positive WKB-wavenumber defined in~(\ref{refl:WKB-0}) and approximate expression for potential $\mathcal{V}(y)$ is given in Eq.~(\ref{refl:Born_approx}) of Appendix~\ref{app.sec:reflection}. The lower limit $y_0\in\left(-\infty ,y_\ast \right)$ of integral in exponent is chosen consistently with the WKB-exponent~(\ref{quasiclassic-sol}) for the incident wave. 

Let us proceed further with the short-wave limit. The developed general approximate analysis in Subsection~\ref{subsec:short-wave-limit} can be applied only partially to describe a wave train absorption near the critical layer. The analysis properly describes the movement of the wave train's center, but fails to describe its width and intensity. Below in Subsection~\ref{subsec:visc_train} we treat the wave equation (\ref{eq:v_invisc}) more rigorously going beyond the envelope approximation (\ref{envelope}) and, in particular, describe the dynamics of the width and intensity of the wave train. 

Here we establish only the movement of a wave train's center $\overline \eta(t)$, considering it localized in the vicinity of some position $\overline \eta_0$ at $t=0$. The current carrier wavenumber is $\overline k_y^\prime = \sigma\beta/(2\overline \eta)$. As the group velocity (\ref{group-velocity-rectilinear}) equals to ${\mathrm v}_{\!g\,}=2 |\Sigma k_x| \overline\eta^2/\beta$, the wave train moves toward the critical layer according to the formula
\begin{equation}\label{eta-bar}
    \overline \eta (t)
    = 
    \frac{\overline \eta_0}{1-2|\Sigma k_x|t\overline \eta_0/\beta}.    
\end{equation}
This leads to the asymptotic approaching as $\overline \eta\propto 1/t$ at large times $t\gg t_\Sigma$, where $t_\Sigma = \oky/|\Sigma k_x |$ is defined via initial carrier wavenumber $\oky=\sigma\beta/(2\overline \eta_0)$.

\section{Absorption in the presence of viscosity} 			\label{sec:viscous_model}
 
The divergent behavior of the velocity field $\bm u$ at the critical layer for a monochromatic wave occurs due to negligence of viscous dissipation effects. To clarify the dissipation of the absorbed wave, we have generalized Eqs.~(\ref{eq:ux},\ref{eq:v_invisc}), restoring finite $\nu$ within the framework of rectilinear model. The procedure is carried out in Appendix \ref{app.sec:viscous_eqs}, and the resulting equations are
\begin{equation}\label{visc:u_eq}
    u
    =
    -\frac{\Delta}{2\Omega k_{{\scriptscriptstyle \perp}}^{2}}
    \big(i\widetilde{\omega}+\nu\Delta \big)v
    +
    \frac{ik_{x}}{k_{{\scriptscriptstyle \perp}}^{2}}
    \partial_{y}\left[\left(1-\frac{U^{\prime}}{\Omega}\right)v\right],
\end{equation}
\begin{multline}\label{visc:v_eq}
    \widetilde{\omega}^{2}\Delta v
    +
    2\Omega\left(2\Omega-U^{\prime}\right)k_{z}^{2}v
    +
    k_{x}U^{\prime\prime}\widetilde{\omega}v
    \\	
    =
    \nu^{2}\Delta^{3}v+i\nu\left(\Delta\widetilde{\omega}
    +
    \widetilde{\omega}\Delta\right)\Delta v
    +
    ik_{x}\nu\Delta U^{\prime\prime}v,
\end{multline}
where we have denoted $\Delta=\partial_y^2-k_{\scriptscriptstyle\perp}^2$ as the Laplacian. The vertical component $w$ can be found from incompressibility condition~(\ref{eq:uz}). Eq.~(\ref{eq:v_invisc}), as inviscid limit $\nu\rightarrow 0$ of~(\ref{visc:v_eq}), is obtained directly; in order to recover~Eq.~(\ref{eq:ux}), one need to take this limit in~(\ref{visc:u_eq}), then use~(\ref{eq:v_invisc}). For internal gravity waves in stratified fluid, similar derivation was done in Ref.~\cite{hazel1967heat&visc} for the two-dimensional problem,  considering both effects of viscosity and heat conductivity.

\subsection{Absorption of monochromatic wave}
\label{subsec:visc_wave}
 
In order to describe  a monochromatic wave inside the viscous inner region around the critical layer, one can simplify the equation (\ref{visc:v_eq}). In the Laplacian $\Delta$, we neglect $k_{\scriptscriptstyle\perp}^2$ compared to the second derivative of $v$ with respect to $\eta=y-y_\ast$, $\partial_\eta^2$ and keep only high-order derivatives of $v$ among viscous terms, similarly to the procedure in~\cite{hazel1967heat&visc}, see~Eq.~(1.6). The result of the simplification in our case can be represented in the following factorized form:
\begin{equation}\label{visc:v-eq_factorized}
\begin{aligned}
    \left(\eta\partial_{\eta}
        +
        i\sigma\eta_{\mathrm{v}}^3\partial_{\eta}^{3}
        -
        \frac{1+ i\beta}{2}\right) \hskip30pt & 
    \\
    \left(\eta\partial_{\eta}
        +
        i\sigma\eta_{\mathrm{v}}^3\partial_{\eta}^{3}
        -
        \frac{1- i\beta}{2}\right)&v
    =
    0.
\end{aligned}
\end{equation}
where we have introduced a positive viscous scale $\eta_{\mathrm{v}}$ according to $\eta_{\mathrm{v}}^3 \equiv \nu/\left|\Sigma k_{x}\right|$ and $\sigma = -\mathrm{sign}\, \Sigma k_x$ is from~(\ref{eq:02}). The brackets in (\ref{visc:v-eq_factorized}) commute with each other. Equation (\ref{visc:v-eq_factorized}) justifies the regularization~(\ref{eq:crit_regularization}) in complex plane $\eta$ for a monochromatic wave motion in inviscid limit, which is implemented now on the viscous scale $\eta_\mathrm{v}$.

\begin{figure}[t!]

\includegraphics[width=\linewidth]{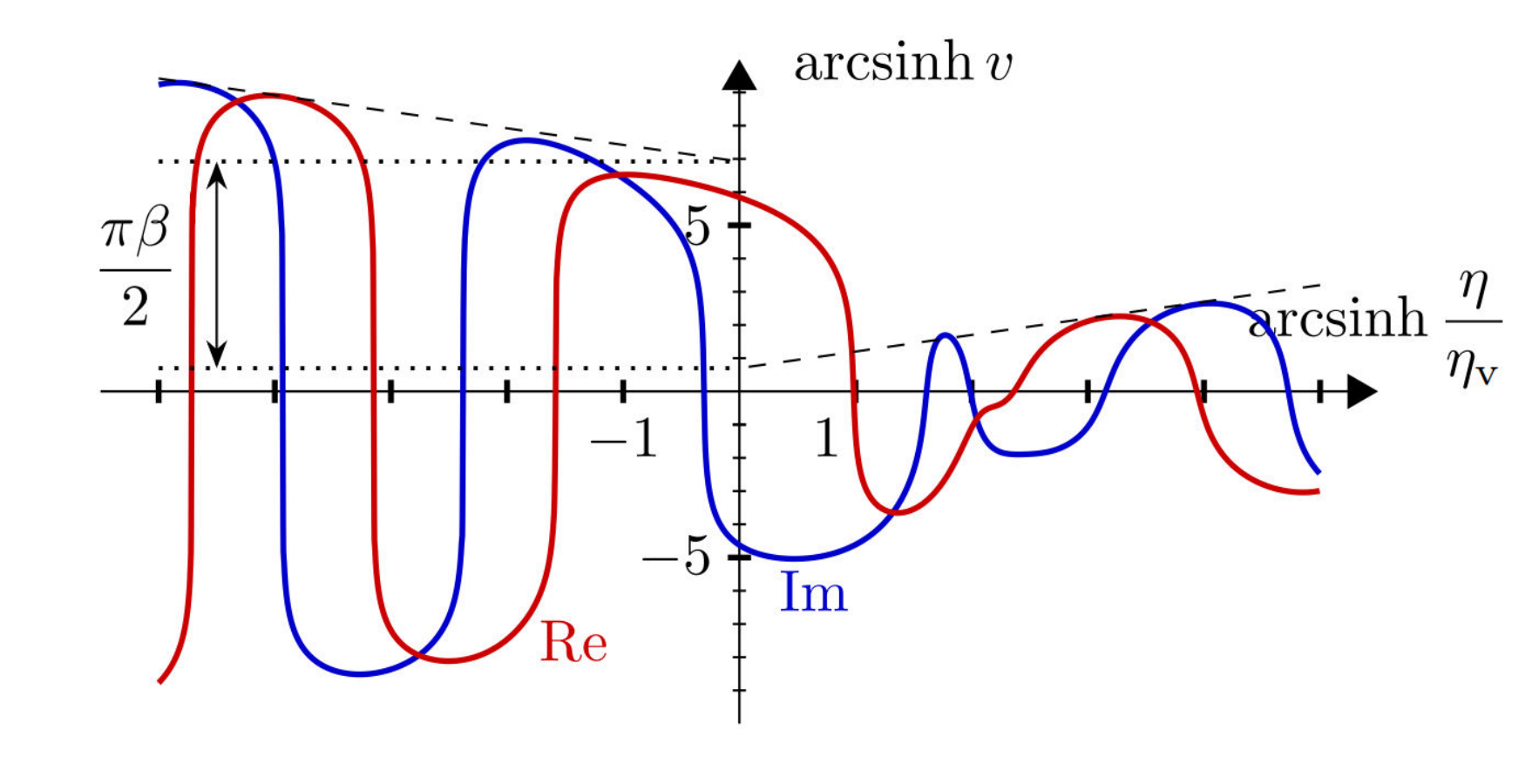}

\caption{Plot of numerical solution of (\ref{visc:v_pm-eq}) for wave $v_{\sigma}$ with $\sigma=1$ passing the viscous inner layer. The solution is obtained by integration of (\ref{visc:v-sol_real}) with $\beta=4$. Red and blue lines correspond to real and imaginary parts of the solution respectively. Dashed lines in both regions $\left|\eta/\eta_{\mathrm v}\gg1\right|$ denote the scaling $|v_+|\propto \sqrt{|\eta|}$ for the absolute value, as per the asymptotic (\ref{eq:02}). The vertical bias between dashed asymptotics corresponds to the absolute value of the transmission coefficient, its numerical value is 2\% less than its theoretical prediction (\ref{eq:transmission}).}
\label{fig:visc_layer-sol}
\end{figure}

The solution of (\ref{visc:v-eq_factorized}) with an asymptotic~(\ref{eq:zero_asympt}) outside the viscous region is a zero mode for one of the two differential operators:
\begin{equation}\label{visc:v_pm-eq}
    \left(\eta\partial_{\eta}
        +
        i\sigma\eta_\mathrm{v}^3\partial_{\eta}^{3}
        -
        \frac{1\pm i\beta}{2}\right)
    v_\pm(\eta)
    =
    0.
\end{equation}
This type of equation arises in context of stability analysis in stratified shear flow as well, see \cite{gage1968stability,maslowe2013study}. The solution of (\ref{visc:v_pm-eq}) can be expressed in terms of generalized Airy class functions described e.g.~in~\cite{drazin_reid_book2004}, where they were introduced for Orr-Sommerfeld equation problem, see Appendix~\ref{app.sec:viscous_eqs} for details. We give here the integral representation formula, where, in contrast with (\ref{visc:v-sol_Airy}), the singularity is integrable on the real axis: 
\begin{equation}\label{visc:v-sol_real}
    \begin{aligned}
    v_{\pm}
    \propto 
    \intop_{-\infty}^\infty
    \frac{\theta(-\sigma k_y)\,\left(i\sigma\eta+\eta_{\mathrm{v}}^3k_y^2\right)d k_y}
        { k_y^{(1\pm i\beta)/2}}
    e^{-i\sigma k_y \eta
        -(\eta_{\mathrm{v}}k_y)^{3}/3}.
    \end{aligned}
\end{equation}
Here $\theta$ is the the Heaviside step function. The choice of the solution as $v_\sigma$ describes the wave travelling to the right through the critical layer, see Eq.~(\ref{eq:02}); its power-law asymptotic takes place in the limit $|\eta/\eta_\mathrm{v}|\gg1$ for (\ref{visc:v-sol_real}). To obtain it, one needs to consider the integrand near the singular point $k_y=0$, where the contour should be tilted at phase $-\sigma\pi/2\cdot\mathop{\mathrm{sign}}\eta$ near the point in complex plane to achieve the fastest convergence of the integral. On the way, one recovers the transmission amplitude~(\ref{eq:transmission}). The characteristic behavior of the solution $v_\sigma (\eta)$ in inner layer is shown in Fig.~\ref{fig:visc_layer-sol}, where -- both for its real and for imaginary parts, -- their inverse hyperbolic sines are plotted against the $\mathrm{arcsinh}(\eta/\eta_{\mathrm{v}})$. This map plots effectively in log-log scale  for values higher than unity, allowing to check the power-law
asymptotics.

\subsection{Absorption of wave train}
\label{subsec:visc_train}

In this Subsection we proceed with the description of a quasi-monochromatic wave train dynamics which was started in Subsection~\ref{subsec:invisc_abs}, including the effects of viscous absorption now. We consider wave train of narrow frequency width $\delta \omega$ around the carrier frequency $\omega$ and determine the distance to the critical layer (for a carrier wave) as $\eta=y-y_\ast(\omega)$. The dynamics of wave train $v_\sigma(t,\eta)$, approaching to the layer from the left, $\eta<0$, is governed by the equation that corresponds to one from the Eq.~(\ref{visc:v_pm-eq}) in the monochromatic case. One replaces $\widetilde \omega \to i\partial_t-\omega +k_x \Sigma \eta$ in $\widetilde\omega\partial_\eta v_\sigma$ term (corresponding to $\Sigma k_x\eta \partial_\eta v_\sigma$  term in~(\ref{visc:v_pm-eq})) and obtains:
\begin{equation}\label{vics:v-train}
   \left(\big(i\partial_{t}-\omega+\Sigma k_{x}\eta\big)\partial_{\eta}-i\nu\partial_{\eta}^{3}-\frac{1+i\sigma\beta}{2}\Sigma k_{x}\right)v_\sigma=0.
\end{equation}
If one uses substitution (\ref{quasiclassic-sol}) with $k_y = \sigma/\rho \eta$ in the short-wave limit $\rho\ll1$, in comparison with (\ref{envelope}) (with zero right-hand side) the result will have additional term $-i\sigma\rho \eta \partial_\eta\partial_t\Phi$, see also~(\ref{visc:envelope}). The term significantly changes dynamics of wave train's form, that will be shown after~(\ref{quadratic}).

Let's introduce the Fourier transform $\widetilde v_\sigma^{\scriptscriptstyle(0)}(k_y)$ of the initial velocity field $v_{\sigma}^{{\scriptscriptstyle (0)}}(\eta)$ at $t=0$. Passing to the Fourier space $\eta \rightarrow k_y$ in Eq.~(\ref{vics:v-train}), one can solve it with the initial condition by the method of characteristics. The expression for $v_\sigma(t,\eta)$ via inverse transform integral is:
\begin{multline}\label{visc:train-sol}
    v_{\sigma}(t,\eta)=e^{-i\omega t}\int\frac{d\ky}{2\pi}e^{ik_y^\prime(t)\eta}\left(\frac{\ky}{k_y^\prime(t)}\right)^{\frac{3+i\sigma\beta}{2}}
    \\
    \times\widetilde v_\sigma^{\scriptscriptstyle(0)}(\ky)\,
    e^{-\nu\intop_{0}^{t}d\tau\left(k_y^\prime(\tau)\right)^{2}}.
\end{multline}
where the integration variable $\ky$ is an initial wavenumber that corresponds to the wavenumber on the characteristic $k_y^\prime(t) = \ky + \Sigma k_x t$. The initial profile $\widetilde v_\sigma^{\scriptscriptstyle(0)}\left(\ky\right)$ for a wave travelling towards the critical layer should satisfy $\sigma \ky<0$, see~(\ref{visc:v-sol_real}). The inequality $\Sigma k_x \ky>0$ means that the singularity $\ky=-\Sigma k_x t$ in (\ref{visc:train-sol}) for $t>0$ is on the half line where $\widetilde v_\sigma^{\scriptscriptstyle(0)} \left(\ky\right)$ equals to zero.

\begin{figure}[t]

\includegraphics[width=\linewidth]{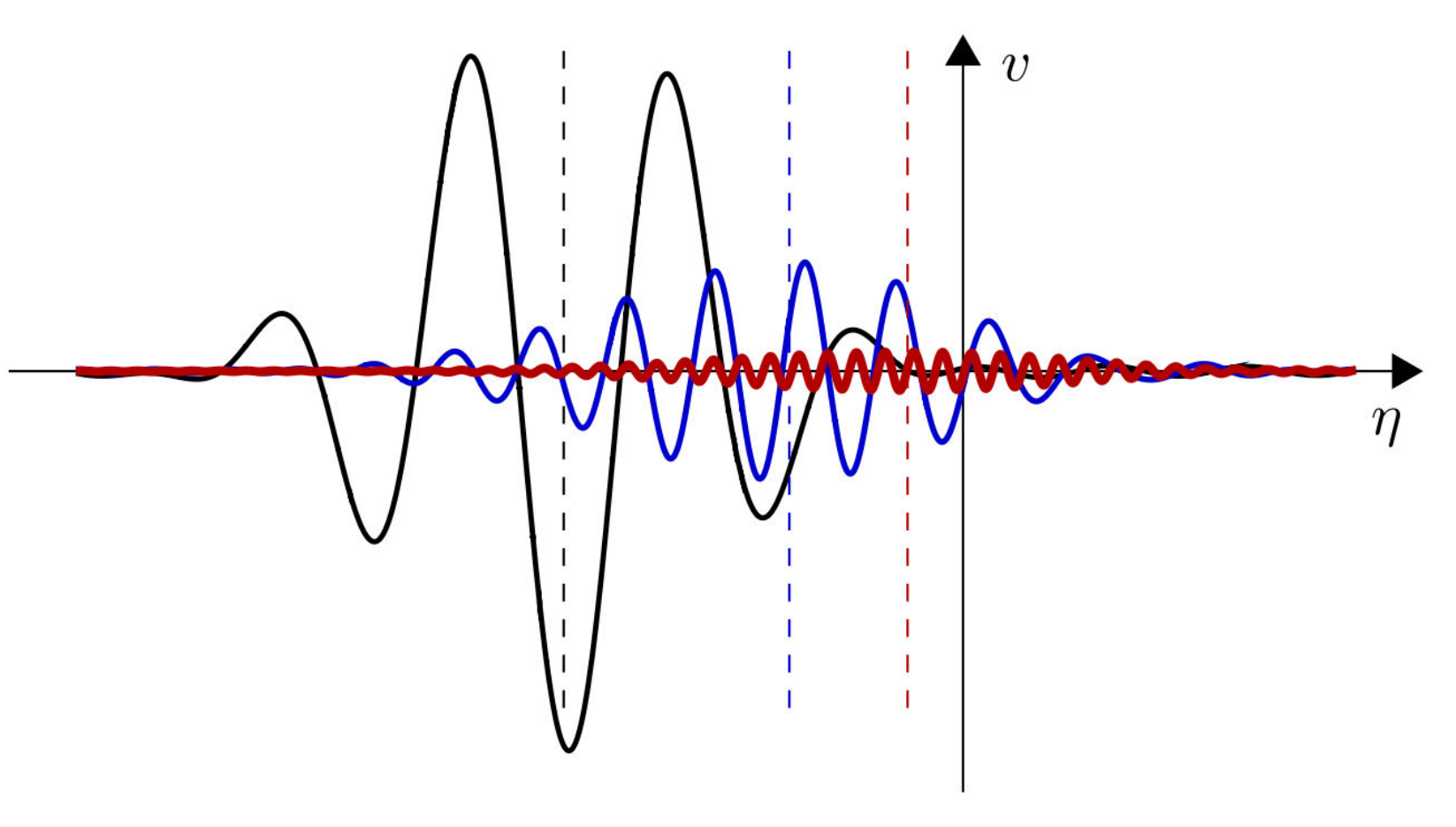}

\caption{Wave train evolution in the vicinity of critical layer in the inviscid limit obtained by numerical integration of (\ref{visc:train-sol}). Taking parameters of width in (\ref{quadratic}) as  $l_0/\overline \eta_0\approx 0.4$ and $\beta=25$, $\mathrm{Re }\:v_+$ function is plotted at times $|\Sigma k_x\overline \eta_0| t = 0, \,15, \, 75$. The train center's position is approaching critical layer and its wavelength is shortening (by $2.2$ and $7$ times respectively, vertical dashed lines represent the center positions, the location of $v$-axis is $\eta=0$) in agreement with (\ref{eta-bar},\ref{quadratic}). The amplitude is decreasing in agreement with $t^{-3/2}$ power-law from Eq.~(\ref{fin-booker}) for times in between blue and red curves (decreased by $3.3$ and $19$ times respectively). Meanwhile, the train's width remains almost unchanged, which follows from~(\ref{quadratic}).}
\label{fig:wave_packets}
\end{figure}

The solution (\ref{visc:train-sol}) corresponds to that obtained in \cite[Eq. (20)]{kolokolov2020structure} for plane wave in a homogeneous shear flow. First, the real part $3/2$ of the first exponent in (\ref{visc:train-sol}) means that absolute value of wave velocity vector's Fourier transform behaves as $|\widetilde {\bm u}(t,\ky)| \propto (\ky/k_y^\prime(t))^{1/2}$ due to (\ref{eq:uz},\ref{eq:ux}), which corresponds to the square root in~\cite[Eq. (20)]{kolokolov2020structure}. Second, the expression in the viscous exponent in (\ref{visc:train-sol}) differs from proper viscous dissipation exponent $-\nu \intop_0^t d\tau \left(\bm k^\prime(\tau)\right)^2$ in the constant shear flow $U_x=-\Sigma y$ by $-\nu t k_{\scriptscriptstyle\perp}^2$, which we have neglected in Eq.~(\ref{visc:v-eq_factorized}). Such dissipation exponent and the corresponding viscous time scale $t_\mathrm{v}=(\nu \Sigma^2 k_x^2)^{-1/3}$ arise essentially for local dynamics in the shear flow case, see~\cite{lebedev2016} for 2D turbulent, \cite{kolokolov2020structure} for 3D model in rotating system or \cite{borovikov1990critical} for internal waves. 

In the short-wave limit, the integral (\ref{visc:train-sol}) can be approximated by means of the saddle-point method based on oscillating exponent $(\ky/k_y^\prime(t))^{i\sigma\beta/2}$. We take it in weak viscosity limit and consider times $t\ll t_{\mathrm{v}}$ first, when the viscosity is negligible. Concerning the initial condition, we presume that the wave train is localized in the vicinity of its center position $\overline \eta_0<0$, which is the point where absolute value $|v_{\sigma}^{{\scriptscriptstyle (0)}}|$ reaches its maximum. The velocity field $v_{\sigma}^{{\scriptscriptstyle (0)}}$ oscillates in space with the carrier wavenumber $\oky$, which is defined as the position of maximum of the absolute value of the Fourier transform $|\widetilde v_{\sigma}^{{\scriptscriptstyle (0)}}|$. The difference $\eta_\ast=\overline\eta_0-\sigma\beta/(2\oky)$ is the position of the critical layer for the carrier wave in the train. Parameters $\oky$, $\overline \eta_0$ and the initial complex width parameter $l_0$ of the wave train in space are connected with the Fourier transform $\widetilde v^{\scriptscriptstyle(0)}_\sigma$ as follows: 
\begin{equation}
    \overline \eta_0 
    =
    i\frac{\partial}{ \partial {k_y}} \ln \widetilde v^{\scriptscriptstyle(0)}_\sigma \big\vert_{k_y = \oky},
    \qquad 
    l_0^2 
    = 
    - \frac{\partial^2}{\partial k_y^2} \ln \widetilde v^{\scriptscriptstyle(0)}_\sigma \big\vert_{k_y = \oky}.
\end{equation}
We assume that initial width $L_0 = |l_0^2|/\sqrt{\mathop{\mathrm{Re}}l_0^2}$ is small compared to the wave train position, $L_0\ll |\overline \eta_0|$. The saddle point $k_y=\oky$ stands for center of the wave train. Position of the center $\overline \eta(t)+\eta_\ast$ agrees with one previously found in the wave train approximation, see~(\ref{eta-bar}) in the end of Subsection~\ref{subsec:invisc_abs}. Assuming the expansion of $\ln \widetilde v^{\scriptscriptstyle(0)}_\sigma$ up to the second order in $\delta k_y = \ky - \oky$ is sufficient to describe the packet near its center, the result of integration in (\ref{visc:train-sol}) is 
\begin{equation}\label{quadratic}
    v_\sigma 
    =
   e^{-i\omega t} \frac{l_0}{l}v_0^{\scriptscriptstyle(0)}
    \left(\frac{\oky}{\overline k_y^\prime}\right)^{\!\frac{3+i\sigma\beta}{2}}
    \exp\left[i\frac{\eta-\eta_\ast}{\rho\overline{\eta}}-\frac{(\delta\eta)^{2}}{2l^{2}}\right],
\end{equation}
where $v_0^{\scriptscriptstyle(0)}$ is initial amplitude in center $\overline \eta_0$, $\overline k_y^\prime = \sigma\beta/(2 \overline \eta)$, $\delta \eta = (\eta-\eta_\ast) - \overline \eta$ is displacement from the wave train center and $l^2 = l_0^2 + i\sigma \rho (\overline \eta_0^2 - \overline \eta^2)$ is square of the wave train's width parameter. The width parameter of the wave train $l$ does not change if its initial value is sufficiently large, $|l_0| \gg \sqrt{\rho} \,\overline \eta_0$. Otherwise, it increases significantly during times $t\sim t_\Sigma$ and then stops to change. The wave train effectively stops to move when its width $L = |l^2|/\sqrt{\mathop{\mathrm{Re}}l^2}$ becomes larger than its position, $L \gg |\overline \eta|$, that occurs at times $t\gg (\eta_0/ L) t_\Sigma$. The established behavior of the wave train's width can not be obtained in framework of the envelope equation (\ref{envelope}), contrary to more rigorous equation~(\ref{vics:v-train}), due to the dispersion term is not accounted in the first one. Despite this discrepancy, the conservation law of wave action (\ref{wave-action}) is fulfilled in the framework of the Equation (\ref{vics:v-train}) as well -- now the following invariant in Fourier space along characteristic $k_y^\prime(t)$ acts as it:
\begin{equation}\label{wave-action1}
    \big|k_y^\prime\big|^3\,\big|\,v_\sigma (k_y^\prime)\big|^2 
    = 
    \mathrm{const}.
\end{equation}
The kinetic energy stored in the wave is proportional to $| k_y^\prime|^2$, and its frequency $\widetilde \omega$ is proportional to $1/| k_y^\prime|$ due to assumed $| k_y^\prime|\gg k_{\scriptscriptstyle\perp}$, so (\ref{wave-action}) and (\ref{wave-action1}) indeed correspond to each other. 

When the wave train stops moving at relatively large times $t\gg (\eta_0/ L) t_\Sigma$, the integral (\ref{visc:train-sol}) can be represented in the form
\begin{equation}\label{fin-booker}
    \begin{aligned}
    v_\sigma 
    = & 
    \frac{e^{-i\omega t +i\Sigma k_{x} t\eta}}
    {2\pi \left(\Sigma k_x t\right)^{(3+i\sigma \beta)/2}}
    \\
    & 
    \times\int d k_y \,
    \ky^{(3+i\sigma \beta)/2}\,
    e^{i\ky \,\eta}\ 
    \widetilde v_\sigma^{\scriptscriptstyle(0)}(\ky).
    \end{aligned}
\end{equation}
The power-law decay in time was previously established in \cite[Eq.~(5.8)]{booker1967}, the Equation (\ref{fin-booker}) adds information about the form of the wave train at the stage.

The carrier wavenumber in (\ref{quadratic},\ref{fin-booker}) continues to increase linearly with time, so eventually the viscosity leads to a fast decay of the wave train amplitude at times $t\gg t_{\mathrm{v}}$, with the exponent $\ln{| v_\sigma|}\sim -t^3/(3t_\mathrm{v}^3)$. The scale~$t_\mathrm{v}$ had been established already as characteristic time it takes for inertial waves to contribute to the Reynolds shear stress $\langle uv\rangle$ mean in~\cite{kolokolov2020structure}.

\section{Conclusions and discussion}
\label{sec:conc}

From the linear approximation of inertial wave dynamics~(\ref{eq:lin-system_cyl}) inside axisymmetric steady geostrophic vortex we have derived the equation~(\ref{eq:v-eq_cyl}) for its radial velocity. If a monochromatic convergent cylindrical wave carries angular momentum with the sign opposite to the sign of the vortex's angular momentum, then it is reflected back at some distance from the vortex axis. In the short-wave limit the reflection takes place at radius that is given by equation (\ref{dispersion03}): the local radial wavenumber $k_r$ reaches zero there. Otherwise, a wave with angular momentum of the same sign may reach its critical layer, where it is partially absorbed by the mean flow. The critical layer inside the vortex is the cylindrical surface $r=r_\ast$ where wave's relative frequency (\ref{eq:Doppler_freq}), subjected to the Doppler shift from the mean flow, reaches zero: $\widetilde\omega (r_\ast)=0$. The layer exists if the mean flow is strong enough. 

Our further analysis was based on the model with rectilinear geometry of the background shear flow. The model allows one to simplify the consideration of the wave behavior in the vicinity of the critical layer or of reflection surface. The critical layer in rectilinear geometry was studied before for inertial waves e.g. in \cite{astoul2021complex} and was well-developed for internal gravity waves problem: basic results in Subsection~\ref{subsec:invisc_abs} are analogous to the theory presented in Ref.~\cite{booker1967}. We considered monochromatic inertial wave as well as wave train dynamics near critical layer in presence of viscosity. First, we derived the general wave equation (\ref{visc:v_eq}) from linear system~(\ref{eq:linear_wave}). Then we showed that it can be simplified to factorized form (\ref{visc:v-eq_factorized}) in the vicinity of the critical layer. The form allows one to analyze separately waves that travel to/from the layer.

The conventional theory of a wave train near the critical layer in the short-wave limit, see e.g. \cite[Eq.~(1.3)]{benney1975nlin} and review~\cite[Eq.~(13)]{maslowe1986}, cannot describe its shape during propagation. In the leading approximation, see Eq.~(\ref{envelope}), wave train travels with group velocity ${\mathrm v}_{\!g\,}$ (\ref{group-velocity-rectilinear}) that describes correctly the movement of the train's center. In the next approximation of the theory, its dynamics would be slightly modified with a relatively weak dispersion which amplitude is proportional to $\partial {\mathrm v}_{\!g\,}/\partial k_y$. If this approach was valid in the case, for a sufficiently wide wave train, the dispersion would be off relevance and the width of the wave train would decrease as $1/t^2$ at large enough times $t$ according to (\ref{envelope}). Our analysis was carried out on the basis of Eq.~(\ref{vics:v-train}), derived from Eq.~(\ref{visc:v_eq}) of viscous linear theory for a wave propagating toward the critical layer and is presented in Subsection~\ref{subsec:visc_train}. It shows that dynamics of train's width contradicts the conventional picture above, remaining constant at large times. This is due to the flow near critical layer is characterized with constant velocity gradient. As applied to $v_\sigma$ solution in Fourier $k_y$-space, see Eq.~(\ref{visc:train-sol}), the flow sets the evolution along characteristics $k^\prime_y(t)=k_y+\Sigma k_x t$, that are parallel to each other, so the spectral width of the wave train is constant in time. If one excludes the monochromatic wave (\ref{eq:zero_asympt}) by substituting $v=\eta^{(1+i\sigma \beta)/2}e^{-i\omega t}a$ (in terms of envelope $\Phi$ (\ref{quasiclassic-sol}) is $\sqrt{\eta} a$), the Eq.~(\ref{vics:v-train}) brings to the following equation for amplitude $a$ in the inviscid limit: 
\begin{equation}\label{visc:envelope}
    \Big(\big(1-i\sigma/\beta\big)\partial_{t}
    +
    {\mathrm v}_{\!g\,}\partial_{\eta}\Big)a
    -
    \frac{2i\sigma\eta}{\beta}\partial_{t}\partial_{\eta}a
    =
    0,
\end{equation}  
compare with (\ref{envelope}), where the group velocity ${\mathrm v}_{\!g\,} = 2 |\Sigma k_x| \eta^2/\beta$. Application of the standard envelope approximation $\partial_t \to -{\mathrm v}_{\!g\,}\partial_\eta$ in the last term is insufficient here. However, it is important to note that the wave action (\ref{wave-action}) is conserved, see~(\ref{wave-action1}), though its concept is originally based on the conventional wave train theory.

Since amplitude of a monochromatic wave increases near its critical layer as $1/\sqrt{\eta}$, one can expect that the nonlinear effects are essential here. For internal gravity waves, the analysis of the nonlinear effects in Ref.~\cite{dornbrack1997lincompare} was provided on the basis of own numerical simulation and the results of experiment~\cite{thorpe1981experiment}. The authors concluded that the linear wave theory was sufficient to describe the wave field, if it is provided with the proper background mean flow profile. In its turn the vortex's mean flow should satisfy the averaged equation that accounts for the divergence $(r^2\Pi_{r \varphi })^\prime/r^2$ of the Reynolds stress~(\ref{eq:momentum_flux}) term from waves. Thus, we see an extension of the present study in applying the developed linear theory of inertial waves to equation for the mean flow that determines vortex flow's velocity profile. Let's make a reservation that this approach implies homogeneity of vertical and azimuthal directions (corresponding to rectilinear $z$- and $x$-axes). This corresponds to considering a wave ensemble, which is statistically homogeneous along these directions. Consideration of a wave train in other case, say when being localized also in x-direction, leads to the nonlinearity violates the homogeneity along the direction, see e.g. numerical study of nonlinear dynamics of internal gravity waves near critical layer~\cite{campbell2003}.

\bibliography{inertial-waves}

\begin{thebibliography}{46}%
\makeatletter
\providecommand \@ifxundefined [1]{%
 \@ifx{#1\undefined}
}%
\providecommand \@ifnum [1]{%
 \ifnum #1\expandafter \@firstoftwo
 \else \expandafter \@secondoftwo
 \fi
}%
\providecommand \@ifx [1]{%
 \ifx #1\expandafter \@firstoftwo
 \else \expandafter \@secondoftwo
 \fi
}%
\providecommand \natexlab [1]{#1}%
\providecommand \enquote  [1]{``#1''}%
\providecommand \bibnamefont  [1]{#1}%
\providecommand \bibfnamefont [1]{#1}%
\providecommand \citenamefont [1]{#1}%
\providecommand \href@noop [0]{\@secondoftwo}%
\providecommand \href [0]{\begingroup \@sanitize@url \@href}%
\providecommand \@href[1]{\@@startlink{#1}\@@href}%
\providecommand \@@href[1]{\endgroup#1\@@endlink}%
\providecommand \@sanitize@url [0]{\catcode `\\12\catcode `\$12\catcode
  `\&12\catcode `\#12\catcode `\^12\catcode `\_12\catcode `\%12\relax}%
\providecommand \@@startlink[1]{}%
\providecommand \@@endlink[0]{}%
\providecommand \url  [0]{\begingroup\@sanitize@url \@url }%
\providecommand \@url [1]{\endgroup\@href {#1}{\urlprefix }}%
\providecommand \urlprefix  [0]{URL }%
\providecommand \Eprint [0]{\href }%
\providecommand \doibase [0]{https://doi.org/}%
\providecommand \selectlanguage [0]{\@gobble}%
\providecommand \bibinfo  [0]{\@secondoftwo}%
\providecommand \bibfield  [0]{\@secondoftwo}%
\providecommand \translation [1]{[#1]}%
\providecommand \BibitemOpen [0]{}%
\providecommand \bibitemStop [0]{}%
\providecommand \bibitemNoStop [0]{.\EOS\space}%
\providecommand \EOS [0]{\spacefactor3000\relax}%
\providecommand \BibitemShut  [1]{\csname bibitem#1\endcsname}%
\let\auto@bib@innerbib\@empty
\bibitem [{\citenamefont {Tumachev}\ \emph {et~al.}(2023)\citenamefont
  {Tumachev}, \citenamefont {Filatov}, \citenamefont {Vergeles},\ and\
  \citenamefont {Levchenko}}]{tumachev2023tworegimes}%
  \BibitemOpen
  \bibfield  {author} {\bibinfo {author} {\bibfnamefont {D.}~\bibnamefont
  {Tumachev}}, \bibinfo {author} {\bibfnamefont {S.}~\bibnamefont {Filatov}},
  \bibinfo {author} {\bibfnamefont {S.}~\bibnamefont {Vergeles}},\ and\
  \bibinfo {author} {\bibfnamefont {A.}~\bibnamefont {Levchenko}},\ }\bibfield
  {title} {\bibinfo {title} {Two dynamical regimes of coherent columnar
  vortices in a rotating fluid},\ }\href
  {https://link.springer.com/article/10.1134/S0021364023602476} {\bibfield
  {journal} {\bibinfo  {journal} {JETP Letters}\ }\textbf {\bibinfo {volume}
  {118}},\ \bibinfo {pages} {426} (\bibinfo {year} {2023})}\BibitemShut
  {NoStop}%
\bibitem [{\citenamefont {Tumachev}\ \emph {et~al.}(2024)\citenamefont
  {Tumachev}, \citenamefont {Levchenko}, \citenamefont {Vergeles},\ and\
  \citenamefont {Filatov}}]{tumachev2024observation}%
  \BibitemOpen
  \bibfield  {author} {\bibinfo {author} {\bibfnamefont {D.~D.}\ \bibnamefont
  {Tumachev}}, \bibinfo {author} {\bibfnamefont {A.~A.}\ \bibnamefont
  {Levchenko}}, \bibinfo {author} {\bibfnamefont {S.~S.}\ \bibnamefont
  {Vergeles}},\ and\ \bibinfo {author} {\bibfnamefont {S.~V.}\ \bibnamefont
  {Filatov}},\ }\bibfield  {title} {\bibinfo {title} {Observation of a large
  stable anticyclone in rotating turbulence},\ }\href
  {https://pubs.aip.org/aip/pof/article-abstract/36/12/126620/3325686/Observation-of-a-large-stable-anticyclone-in}
  {\bibfield  {journal} {\bibinfo  {journal} {Physics of Fluids}\ }\textbf
  {\bibinfo {volume} {36}},\ \bibinfo {pages} {126620} (\bibinfo {year}
  {2024})}\BibitemShut {NoStop}%
\bibitem [{\citenamefont {Greenspan}(1990)}]{greenspan1968rotating}%
  \BibitemOpen
  \bibfield  {author} {\bibinfo {author} {\bibfnamefont {H.}~\bibnamefont
  {Greenspan}},\ }\href {https://books.google.com/books?id=2O5QAAAAMAAJ} {\emph
  {\bibinfo {title} {The Theory of Rotating Fluids}}},\ Cambridge Monographs on
  Mechanics and Applied Mathematics\ (\bibinfo  {publisher} {Breukelen Press},\
  \bibinfo {year} {1990})\BibitemShut {NoStop}%
\bibitem [{\citenamefont {Galtier}(2003)}]{galtier2003weak}%
  \BibitemOpen
  \bibfield  {author} {\bibinfo {author} {\bibfnamefont {S.}~\bibnamefont
  {Galtier}},\ }\bibfield  {title} {\bibinfo {title} {Weak inertial-wave
  turbulence theory},\ }\href
  {https://journals.aps.org/pre/abstract/10.1103/PhysRevE.68.015301} {\bibfield
   {journal} {\bibinfo  {journal} {Physical Review E}\ }\textbf {\bibinfo
  {volume} {68}},\ \bibinfo {pages} {015301} (\bibinfo {year}
  {2003})}\BibitemShut {NoStop}%
\bibitem [{\citenamefont {Gelash}\ \emph {et~al.}(2017)\citenamefont {Gelash},
  \citenamefont {L’vov},\ and\ \citenamefont
  {Zakharov}}]{gelash2017complete}%
  \BibitemOpen
  \bibfield  {author} {\bibinfo {author} {\bibfnamefont {A.}~\bibnamefont
  {Gelash}}, \bibinfo {author} {\bibfnamefont {V.}~\bibnamefont {L’vov}},\
  and\ \bibinfo {author} {\bibfnamefont {V.}~\bibnamefont {Zakharov}},\
  }\bibfield  {title} {\bibinfo {title} {Complete hamiltonian formalism for
  inertial waves in rotating fluids},\ }\href
  {https://www.cambridge.org/core/journals/journal-of-fluid-mechanics/article/abs/complete-hamiltonian-formalism-for-inertial-waves-in-rotating-fluids/EAF73F46E26C1DE11F6A34CDA568A246}
  {\bibfield  {journal} {\bibinfo  {journal} {Journal of Fluid Mechanics}\
  }\textbf {\bibinfo {volume} {831}},\ \bibinfo {pages} {128} (\bibinfo {year}
  {2017})}\BibitemShut {NoStop}%
\bibitem [{\citenamefont {Monsalve}\ \emph {et~al.}(2020)\citenamefont
  {Monsalve}, \citenamefont {Brunet}, \citenamefont {Gallet},\ and\
  \citenamefont {Cortet}}]{monsalve2020quantitative}%
  \BibitemOpen
  \bibfield  {author} {\bibinfo {author} {\bibfnamefont {E.}~\bibnamefont
  {Monsalve}}, \bibinfo {author} {\bibfnamefont {M.}~\bibnamefont {Brunet}},
  \bibinfo {author} {\bibfnamefont {B.}~\bibnamefont {Gallet}},\ and\ \bibinfo
  {author} {\bibfnamefont {P.-P.}\ \bibnamefont {Cortet}},\ }\bibfield  {title}
  {\bibinfo {title} {Quantitative experimental observation of weak
  inertial-wave turbulence},\ }\href
  {https://journals.aps.org/prl/abstract/10.1103/PhysRevLett.125.254502}
  {\bibfield  {journal} {\bibinfo  {journal} {Physical Review Letters}\
  }\textbf {\bibinfo {volume} {125}},\ \bibinfo {pages} {254502} (\bibinfo
  {year} {2020})}\BibitemShut {NoStop}%
\bibitem [{\citenamefont {Davidson}(2013)}]{davidson2013turbulence}%
  \BibitemOpen
  \bibfield  {author} {\bibinfo {author} {\bibfnamefont {P.~A.}\ \bibnamefont
  {Davidson}},\ }\href@noop {} {\emph {\bibinfo {title} {Turbulence in
  rotating, stratified and electrically conducting fluids}}}\ (\bibinfo
  {publisher} {Cambridge University Press},\ \bibinfo {address} {Cambridge},\
  \bibinfo {year} {2013})\BibitemShut {NoStop}%
\bibitem [{\citenamefont {Gallet}(2015)}]{gallet2015exact}%
  \BibitemOpen
  \bibfield  {author} {\bibinfo {author} {\bibfnamefont {B.}~\bibnamefont
  {Gallet}},\ }\bibfield  {title} {\bibinfo {title} {Exact
  two-dimensionalization of rapidly rotating large-reynolds-number flows},\
  }\href@noop {} {\bibfield  {journal} {\bibinfo  {journal} {Journal of Fluid
  Mechanics}\ }\textbf {\bibinfo {volume} {783}},\ \bibinfo {pages} {412}
  (\bibinfo {year} {2015})}\BibitemShut {NoStop}%
\bibitem [{\citenamefont {Godeferd}\ and\ \citenamefont
  {Moisy}(2015)}]{godeferd2015structure}%
  \BibitemOpen
  \bibfield  {author} {\bibinfo {author} {\bibfnamefont {F.~S.}\ \bibnamefont
  {Godeferd}}\ and\ \bibinfo {author} {\bibfnamefont {F.}~\bibnamefont
  {Moisy}},\ }\bibfield  {title} {\bibinfo {title} {Structure and dynamics of
  rotating turbulence: a review of recent experimental and numerical results},\
  }\href
  {https://asmedigitalcollection.asme.org/appliedmechanicsreviews/article-abstract/67/3/030802/370036/Structure-and-Dynamics-of-Rotating-Turbulence-A}
  {\bibfield  {journal} {\bibinfo  {journal} {Applied Mechanics Reviews}\
  }\textbf {\bibinfo {volume} {67}},\ \bibinfo {pages} {030802} (\bibinfo
  {year} {2015})}\BibitemShut {NoStop}%
\bibitem [{\citenamefont {McEwan}(1976)}]{mcewan1976angular}%
  \BibitemOpen
  \bibfield  {author} {\bibinfo {author} {\bibfnamefont {A.}~\bibnamefont
  {McEwan}},\ }\bibfield  {title} {\bibinfo {title} {Angular momentum diffusion
  and the initiation of cyclones},\ }\href@noop {} {\bibfield  {journal}
  {\bibinfo  {journal} {Nature}\ }\textbf {\bibinfo {volume} {260}},\ \bibinfo
  {pages} {126} (\bibinfo {year} {1976})}\BibitemShut {NoStop}%
\bibitem [{\citenamefont {Ruppert-Felsot}\ \emph {et~al.}(2005)\citenamefont
  {Ruppert-Felsot}, \citenamefont {Praud}, \citenamefont {Sharon},\ and\
  \citenamefont {Swinney}}]{ruppert2005extraction}%
  \BibitemOpen
  \bibfield  {author} {\bibinfo {author} {\bibfnamefont {J.~E.}\ \bibnamefont
  {Ruppert-Felsot}}, \bibinfo {author} {\bibfnamefont {O.}~\bibnamefont
  {Praud}}, \bibinfo {author} {\bibfnamefont {E.}~\bibnamefont {Sharon}},\ and\
  \bibinfo {author} {\bibfnamefont {H.~L.}\ \bibnamefont {Swinney}},\
  }\bibfield  {title} {\bibinfo {title} {Extraction of coherent structures in a
  rotating turbulent flow experiment},\ }\href
  {https://journals.aps.org/pre/abstract/10.1103/PhysRevE.72.016311} {\bibfield
   {journal} {\bibinfo  {journal} {Physical Review E}\ }\textbf {\bibinfo
  {volume} {72}},\ \bibinfo {pages} {016311} (\bibinfo {year}
  {2005})}\BibitemShut {NoStop}%
\bibitem [{\citenamefont {Hopfinger}\ \emph {et~al.}(1982)\citenamefont
  {Hopfinger}, \citenamefont {Browand},\ and\ \citenamefont
  {Gagne}}]{hopfinger1982turbulence}%
  \BibitemOpen
  \bibfield  {author} {\bibinfo {author} {\bibfnamefont {E.}~\bibnamefont
  {Hopfinger}}, \bibinfo {author} {\bibfnamefont {F.}~\bibnamefont {Browand}},\
  and\ \bibinfo {author} {\bibfnamefont {Y.}~\bibnamefont {Gagne}},\ }\bibfield
   {title} {\bibinfo {title} {Turbulence and waves in a rotating tank},\
  }\href@noop {} {\bibfield  {journal} {\bibinfo  {journal} {Journal of Fluid
  Mechanics}\ }\textbf {\bibinfo {volume} {125}},\ \bibinfo {pages} {505}
  (\bibinfo {year} {1982})}\BibitemShut {NoStop}%
\bibitem [{\citenamefont {Gallet}\ \emph {et~al.}(2014)\citenamefont {Gallet},
  \citenamefont {Campagne}, \citenamefont {Cortet},\ and\ \citenamefont
  {Moisy}}]{gallet2014scale}%
  \BibitemOpen
  \bibfield  {author} {\bibinfo {author} {\bibfnamefont {B.}~\bibnamefont
  {Gallet}}, \bibinfo {author} {\bibfnamefont {A.}~\bibnamefont {Campagne}},
  \bibinfo {author} {\bibfnamefont {P.-P.}\ \bibnamefont {Cortet}},\ and\
  \bibinfo {author} {\bibfnamefont {F.}~\bibnamefont {Moisy}},\ }\bibfield
  {title} {\bibinfo {title} {Scale-dependent cyclone-anticyclone asymmetry in a
  forced rotating turbulence experiment},\ }\href
  {https://pubs.aip.org/aip/pof/article-abstract/26/3/035108/258648/Scale-dependent-cyclone-anticyclone-asymmetry-in-a}
  {\bibfield  {journal} {\bibinfo  {journal} {Physics of Fluids}\ }\textbf
  {\bibinfo {volume} {26}},\ \bibinfo {pages} {035108} (\bibinfo {year}
  {2014})}\BibitemShut {NoStop}%
\bibitem [{\citenamefont {Kraichnan}(1973)}]{kraichnan1973helical}%
  \BibitemOpen
  \bibfield  {author} {\bibinfo {author} {\bibfnamefont {R.~H.}\ \bibnamefont
  {Kraichnan}},\ }\bibfield  {title} {\bibinfo {title} {Helical turbulence and
  absolute equilibrium},\ }\href@noop {} {\bibfield  {journal} {\bibinfo
  {journal} {Journal of Fluid Mechanics}\ }\textbf {\bibinfo {volume} {59}},\
  \bibinfo {pages} {745} (\bibinfo {year} {1973})}\BibitemShut {NoStop}%
\bibitem [{\citenamefont {Booker}\ and\ \citenamefont
  {Bretherton}(1967)}]{booker1967}%
  \BibitemOpen
  \bibfield  {author} {\bibinfo {author} {\bibfnamefont {J.~R.}\ \bibnamefont
  {Booker}}\ and\ \bibinfo {author} {\bibfnamefont {F.~P.}\ \bibnamefont
  {Bretherton}},\ }\bibfield  {title} {\bibinfo {title} {The critical layer for
  internal gravity waves in a shear flow},\ }\href
  {https://www.cambridge.org/core/journals/journal-of-fluid-mechanics/article/critical-layer-for-internal-gravity-waves-in-a-shear-flow/8D72D50D8C40AFCC668E7047B2300EEF}
  {\bibfield  {journal} {\bibinfo  {journal} {Journal of Fluid Mechanics}\
  }\textbf {\bibinfo {volume} {27}},\ \bibinfo {pages} {513} (\bibinfo {year}
  {1967})}\BibitemShut {NoStop}%
\bibitem [{\citenamefont {Baruteau}\ and\ \citenamefont
  {Rieutord}(2013)}]{baruteau2013inertial}%
  \BibitemOpen
  \bibfield  {author} {\bibinfo {author} {\bibfnamefont {C.}~\bibnamefont
  {Baruteau}}\ and\ \bibinfo {author} {\bibfnamefont {M.}~\bibnamefont
  {Rieutord}},\ }\bibfield  {title} {\bibinfo {title} {Inertial waves in a
  differentially rotating spherical shell},\ }\href
  {https://www.cambridge.org/core/journals/journal-of-fluid-mechanics/article/abs/inertial-waves-in-a-differentially-rotating-spherical-shell/055824F6C3B3DD6DAB900E15044CA76D}
  {\bibfield  {journal} {\bibinfo  {journal} {Journal of Fluid Mechanics}\
  }\textbf {\bibinfo {volume} {719}},\ \bibinfo {pages} {47} (\bibinfo {year}
  {2013})}\BibitemShut {NoStop}%
\bibitem [{\citenamefont {Astoul}\ \emph {et~al.}(2021)\citenamefont {Astoul},
  \citenamefont {Park}, \citenamefont {Mathis}, \citenamefont {Baruteau},\ and\
  \citenamefont {Gallet}}]{astoul2021complex}%
  \BibitemOpen
  \bibfield  {author} {\bibinfo {author} {\bibfnamefont {A.}~\bibnamefont
  {Astoul}}, \bibinfo {author} {\bibfnamefont {J.}~\bibnamefont {Park}},
  \bibinfo {author} {\bibfnamefont {S.}~\bibnamefont {Mathis}}, \bibinfo
  {author} {\bibfnamefont {C.}~\bibnamefont {Baruteau}},\ and\ \bibinfo
  {author} {\bibfnamefont {F.}~\bibnamefont {Gallet}},\ }\bibfield  {title}
  {\bibinfo {title} {{The complex interplay between tidal inertial waves and
  zonal flows in differentially rotating stellar and planetary convective
  regions. I. Free waves}},\ }\href
  {https://www.aanda.org/articles/aa/abs/2021/03/aa39148-20/aa39148-20.html}
  {\bibfield  {journal} {\bibinfo  {journal} {Astronomy \& Astrophysics}\
  }\textbf {\bibinfo {volume} {647}},\ \bibinfo {pages} {A144} (\bibinfo {year}
  {2021})}\BibitemShut {NoStop}%
\bibitem [{\citenamefont {Kolokolov}\ \emph {et~al.}(2020)\citenamefont
  {Kolokolov}, \citenamefont {Ogorodnikov},\ and\ \citenamefont
  {Vergeles}}]{kolokolov2020structure}%
  \BibitemOpen
  \bibfield  {author} {\bibinfo {author} {\bibfnamefont {I.}~\bibnamefont
  {Kolokolov}}, \bibinfo {author} {\bibfnamefont {L.}~\bibnamefont
  {Ogorodnikov}},\ and\ \bibinfo {author} {\bibfnamefont {S.}~\bibnamefont
  {Vergeles}},\ }\bibfield  {title} {\bibinfo {title} {Structure of coherent
  columnar vortices in three-dimensional rotating turbulent flow},\ }\href
  {https://journals.aps.org/prfluids/abstract/10.1103/PhysRevFluids.5.034604}
  {\bibfield  {journal} {\bibinfo  {journal} {Physical Review Fluids}\ }\textbf
  {\bibinfo {volume} {5}},\ \bibinfo {pages} {034604} (\bibinfo {year}
  {2020})}\BibitemShut {NoStop}%
\bibitem [{\citenamefont {Parfenyev}\ and\ \citenamefont
  {Vergeles}(2021)}]{parfenyev2021influence}%
  \BibitemOpen
  \bibfield  {author} {\bibinfo {author} {\bibfnamefont {V.~M.}\ \bibnamefont
  {Parfenyev}}\ and\ \bibinfo {author} {\bibfnamefont {S.~S.}\ \bibnamefont
  {Vergeles}},\ }\bibfield  {title} {\bibinfo {title} {{Influence of Ekman
  friction on the velocity profile of a coherent vortex in a three-dimensional
  rotating turbulent flow}},\ }\href
  {https://pubs.aip.org/aip/pof/article-abstract/33/11/115128/1063802/Influence-of-Ekman-friction-on-the-velocity}
  {\bibfield  {journal} {\bibinfo  {journal} {Physics of Fluids}\ }\textbf
  {\bibinfo {volume} {33}},\ \bibinfo {pages} {115128} (\bibinfo {year}
  {2021})}\BibitemShut {NoStop}%
\bibitem [{\citenamefont {Parfenyev}\ \emph {et~al.}(2021)\citenamefont
  {Parfenyev}, \citenamefont {Vointsev}, \citenamefont {Skoba},\ and\
  \citenamefont {Vergeles}}]{parfenyev2021velocity}%
  \BibitemOpen
  \bibfield  {author} {\bibinfo {author} {\bibfnamefont {V.~M.}\ \bibnamefont
  {Parfenyev}}, \bibinfo {author} {\bibfnamefont {I.~A.}\ \bibnamefont
  {Vointsev}}, \bibinfo {author} {\bibfnamefont {A.~O.}\ \bibnamefont
  {Skoba}},\ and\ \bibinfo {author} {\bibfnamefont {S.~S.}\ \bibnamefont
  {Vergeles}},\ }\bibfield  {title} {\bibinfo {title} {Velocity profiles of
  cyclones and anticyclones in a rotating turbulent flow},\ }\href
  {https://pubs.aip.org/aip/pof/article-abstract/33/6/065117/1065707/Velocity-profiles-of-cyclones-and-anticyclones-in}
  {\bibfield  {journal} {\bibinfo  {journal} {Physics of Fluids}\ }\textbf
  {\bibinfo {volume} {33}},\ \bibinfo {pages} {065117} (\bibinfo {year}
  {2021})}\BibitemShut {NoStop}%
\bibitem [{\citenamefont {Sutherland}(2010)}]{sutherland2010internal}%
  \BibitemOpen
  \bibfield  {author} {\bibinfo {author} {\bibfnamefont {B.}~\bibnamefont
  {Sutherland}},\ }\href {https://books.google.ru/books?id=Scyy92I7NLAC} {\emph
  {\bibinfo {title} {Internal Gravity Waves}}}\ (\bibinfo  {publisher}
  {Cambridge University Press},\ \bibinfo {year} {2010})\BibitemShut {NoStop}%
\bibitem [{\citenamefont {Bretherton}\ and\ \citenamefont
  {Garrett}(1968)}]{bretherton1968wavetrains}%
  \BibitemOpen
  \bibfield  {author} {\bibinfo {author} {\bibfnamefont {F.~P.}\ \bibnamefont
  {Bretherton}}\ and\ \bibinfo {author} {\bibfnamefont {C.~J.~R.}\ \bibnamefont
  {Garrett}},\ }\bibfield  {title} {\bibinfo {title} {Wavetrains in
  inhomogeneous moving media},\ }\href
  {https://royalsocietypublishing.org/doi/abs/10.1098/rspa.1968.0034}
  {\bibfield  {journal} {\bibinfo  {journal} {Proceedings of the Royal Society
  of London. Series A. Mathematical and Physical Sciences}\ }\textbf {\bibinfo
  {volume} {302}},\ \bibinfo {pages} {529} (\bibinfo {year}
  {1968})}\BibitemShut {NoStop}%
\bibitem [{\citenamefont {Andrews}\ and\ \citenamefont
  {McIntyre}(1978)}]{andrews1978wave}%
  \BibitemOpen
  \bibfield  {author} {\bibinfo {author} {\bibfnamefont {D.~G.}\ \bibnamefont
  {Andrews}}\ and\ \bibinfo {author} {\bibfnamefont {M.}~\bibnamefont
  {McIntyre}},\ }\bibfield  {title} {\bibinfo {title} {On wave-action and its
  relatives},\ }\href@noop {} {\bibfield  {journal} {\bibinfo  {journal}
  {Journal of Fluid Mechanics}\ }\textbf {\bibinfo {volume} {89}},\ \bibinfo
  {pages} {647} (\bibinfo {year} {1978})}\BibitemShut {NoStop}%
\bibitem [{\citenamefont {Hazel}(1967)}]{hazel1967heat&visc}%
  \BibitemOpen
  \bibfield  {author} {\bibinfo {author} {\bibfnamefont {P.}~\bibnamefont
  {Hazel}},\ }\bibfield  {title} {\bibinfo {title} {The effect of viscosity and
  heat conduction on internal gravity waves at a critical level},\ }\href
  {https://doi.org/10.1017/S0022112067001752} {\bibfield  {journal} {\bibinfo
  {journal} {Journal of Fluid Mechanics}\ }\textbf {\bibinfo {volume} {30}},\
  \bibinfo {pages} {775–783} (\bibinfo {year} {1967})}\BibitemShut {NoStop}%
\bibitem [{\citenamefont {Salhi}\ and\ \citenamefont
  {Cambon}(1997)}]{salhi1997analysis}%
  \BibitemOpen
  \bibfield  {author} {\bibinfo {author} {\bibfnamefont {A.}~\bibnamefont
  {Salhi}}\ and\ \bibinfo {author} {\bibfnamefont {C.}~\bibnamefont {Cambon}},\
  }\bibfield  {title} {\bibinfo {title} {An analysis of rotating shear flow
  using linear theory and dns and les results},\ }\href
  {https://www.cambridge.org/core/journals/journal-of-fluid-mechanics/article/an-analysis-of-rotating-shear-flow-using-linear-theory-and-dns-and-les-results/E2CA629A7B6D6D32989BD346017D500F}
  {\bibfield  {journal} {\bibinfo  {journal} {Journal of Fluid Mechanics}\
  }\textbf {\bibinfo {volume} {347}},\ \bibinfo {pages} {171} (\bibinfo {year}
  {1997})}\BibitemShut {NoStop}%
\bibitem [{\citenamefont {Drazin}\ and\ \citenamefont
  {Reid}(2004)}]{drazin_reid_book2004}%
  \BibitemOpen
  \bibfield  {author} {\bibinfo {author} {\bibfnamefont {P.~G.}\ \bibnamefont
  {Drazin}}\ and\ \bibinfo {author} {\bibfnamefont {W.~H.}\ \bibnamefont
  {Reid}},\ }\href@noop {} {\emph {\bibinfo {title} {Hydrodynamic
  Stability}}},\ \bibinfo {edition} {2nd}\ ed.,\ Cambridge Mathematical
  Library\ (\bibinfo  {publisher} {Cambridge University Press},\ \bibinfo
  {year} {2004})\BibitemShut {NoStop}%
\bibitem [{\citenamefont {Herring}(1974)}]{herring1974}%
  \BibitemOpen
  \bibfield  {author} {\bibinfo {author} {\bibfnamefont {J.~R.}\ \bibnamefont
  {Herring}},\ }\bibfield  {title} {\bibinfo {title} {Approach of axisymmetric
  turbulence to isotropy},\ }\href {https://doi.org/10.1063/1.1694822}
  {\bibfield  {journal} {\bibinfo  {journal} {The Physics of Fluids}\ }\textbf
  {\bibinfo {volume} {17}},\ \bibinfo {pages} {859} (\bibinfo {year}
  {1974})}\BibitemShut {NoStop}%
\bibitem [{\citenamefont {Sagaut}\ and\ \citenamefont
  {Cambon}(2008)}]{sagaut2008homogeneous}%
  \BibitemOpen
  \bibfield  {author} {\bibinfo {author} {\bibfnamefont {P.}~\bibnamefont
  {Sagaut}}\ and\ \bibinfo {author} {\bibfnamefont {C.}~\bibnamefont
  {Cambon}},\ }\href {https://books.google.com/books?id=zo1QIn5lS5IC} {\emph
  {\bibinfo {title} {Homogeneous Turbulence Dynamics}}}\ (\bibinfo  {publisher}
  {Cambridge University Press},\ \bibinfo {year} {2008})\BibitemShut {NoStop}%
\bibitem [{\citenamefont {Miles}(1961)}]{miles1961}%
  \BibitemOpen
  \bibfield  {author} {\bibinfo {author} {\bibfnamefont {J.~W.}\ \bibnamefont
  {Miles}},\ }\bibfield  {title} {\bibinfo {title} {On the stability of
  heterogeneous shear flows},\ }\href
  {https://doi.org/10.1017/S0022112061000305} {\bibfield  {journal} {\bibinfo
  {journal} {Journal of Fluid Mechanics}\ }\textbf {\bibinfo {volume} {10}},\
  \bibinfo {pages} {496} (\bibinfo {year} {1961})}\BibitemShut {NoStop}%
\bibitem [{\citenamefont {Howard}(1961)}]{howard1961}%
  \BibitemOpen
  \bibfield  {author} {\bibinfo {author} {\bibfnamefont {L.~N.}\ \bibnamefont
  {Howard}},\ }\bibfield  {title} {\bibinfo {title} {Note on a paper of john w.
  miles},\ }\href {https://doi.org/10.1017/S0022112061000317} {\bibfield
  {journal} {\bibinfo  {journal} {Journal of Fluid Mechanics}\ }\textbf
  {\bibinfo {volume} {10}},\ \bibinfo {pages} {509} (\bibinfo {year}
  {1961})}\BibitemShut {NoStop}%
\bibitem [{\citenamefont {Berry}\ and\ \citenamefont
  {Mount}(1972)}]{berry1972WKB}%
  \BibitemOpen
  \bibfield  {author} {\bibinfo {author} {\bibfnamefont {M.~V.}\ \bibnamefont
  {Berry}}\ and\ \bibinfo {author} {\bibfnamefont {K.~E.}\ \bibnamefont
  {Mount}},\ }\bibfield  {title} {\bibinfo {title} {Semiclassical
  approximations in wave mechanics},\ }\href
  {https://doi.org/10.1088/0034-4885/35/1/306} {\bibfield  {journal} {\bibinfo
  {journal} {Reports on Progress in Physics}\ }\textbf {\bibinfo {volume}
  {35}},\ \bibinfo {pages} {315} (\bibinfo {year} {1972})}\BibitemShut
  {NoStop}%
\bibitem [{\citenamefont {Badulin}\ \emph {et~al.}(1985)\citenamefont
  {Badulin}, \citenamefont {Shrira},\ and\ \citenamefont
  {Tsimring}}]{badulin1985trapping}%
  \BibitemOpen
  \bibfield  {author} {\bibinfo {author} {\bibfnamefont {S.}~\bibnamefont
  {Badulin}}, \bibinfo {author} {\bibfnamefont {V.}~\bibnamefont {Shrira}},\
  and\ \bibinfo {author} {\bibfnamefont {L.~S.}\ \bibnamefont {Tsimring}},\
  }\bibfield  {title} {\bibinfo {title} {The trapping and vertical focusing of
  internal waves in a pycnocline due to the horizontal inhomogeneities of
  density and currents},\ }\href@noop {} {\bibfield  {journal} {\bibinfo
  {journal} {Journal of Fluid Mechanics}\ }\textbf {\bibinfo {volume} {158}},\
  \bibinfo {pages} {199} (\bibinfo {year} {1985})}\BibitemShut {NoStop}%
\bibitem [{\citenamefont {Badulin}\ and\ \citenamefont
  {Shrira}(1993)}]{badulin1993irreversibility}%
  \BibitemOpen
  \bibfield  {author} {\bibinfo {author} {\bibfnamefont {S.~I.}\ \bibnamefont
  {Badulin}}\ and\ \bibinfo {author} {\bibfnamefont {V.~I.}\ \bibnamefont
  {Shrira}},\ }\bibfield  {title} {\bibinfo {title} {On the irreversibility of
  internal-wave dynamics due to wave trapping by mean flow inhomogeneities.
  part 1. local analysis},\ }\href@noop {} {\bibfield  {journal} {\bibinfo
  {journal} {Journal of Fluid Mechanics}\ }\textbf {\bibinfo {volume} {251}},\
  \bibinfo {pages} {21} (\bibinfo {year} {1993})}\BibitemShut {NoStop}%
\bibitem [{\citenamefont {Miles}(1967)}]{miles1967}%
  \BibitemOpen
  \bibfield  {author} {\bibinfo {author} {\bibfnamefont {J.~W.}\ \bibnamefont
  {Miles}},\ }\bibfield  {title} {\bibinfo {title} {Internal waves in a
  continuously stratified atmosphere or ocean},\ }\href
  {https://api.semanticscholar.org/CorpusID:122632267} {\bibfield  {journal}
  {\bibinfo  {journal} {Journal of Fluid Mechanics}\ }\textbf {\bibinfo
  {volume} {28}},\ \bibinfo {pages} {305} (\bibinfo {year} {1967})}\BibitemShut
  {NoStop}%
\bibitem [{\citenamefont {Drazin}\ \emph {et~al.}(1979)\citenamefont {Drazin},
  \citenamefont {Zaturska},\ and\ \citenamefont {Banks}}]{drazin1979}%
  \BibitemOpen
  \bibfield  {author} {\bibinfo {author} {\bibfnamefont {P.}~\bibnamefont
  {Drazin}}, \bibinfo {author} {\bibfnamefont {M.}~\bibnamefont {Zaturska}},\
  and\ \bibinfo {author} {\bibfnamefont {W.}~\bibnamefont {Banks}},\ }\bibfield
   {title} {\bibinfo {title} {On the normal modes of parallel flow of inviscid
  stratified fluid. part 2. unbounded flow with propagation at infinity},\
  }\href
  {https://www.cambridge.org/core/journals/journal-of-fluid-mechanics/article/on-the-normal-modes-of-parallel-flow-of-inviscid-stratified-fluid-part-2-unbounded-flow-with-propagation-at-infinity/CDFEB06C69EEB79584577017A0AFBE6A}
  {\bibfield  {journal} {\bibinfo  {journal} {Journal of Fluid Mechanics}\
  }\textbf {\bibinfo {volume} {95}},\ \bibinfo {pages} {681} (\bibinfo {year}
  {1979})}\BibitemShut {NoStop}%
\bibitem [{\citenamefont {Duin}\ and\ \citenamefont {Kelder}(1982)}]{duin1982}%
  \BibitemOpen
  \bibfield  {author} {\bibinfo {author} {\bibfnamefont {C.~A.~V.}\
  \bibnamefont {Duin}}\ and\ \bibinfo {author} {\bibfnamefont {H.}~\bibnamefont
  {Kelder}},\ }\bibfield  {title} {\bibinfo {title} {Reflection properties of
  internal gravity waves incident upon a hyperbolic tangent shear layer},\
  }\href {https://doi.org/10.1017/S0022112082002870} {\bibfield  {journal}
  {\bibinfo  {journal} {Journal of Fluid Mechanics}\ }\textbf {\bibinfo
  {volume} {120}},\ \bibinfo {pages} {505} (\bibinfo {year}
  {1982})}\BibitemShut {NoStop}%
\bibitem [{\citenamefont {Gage}\ and\ \citenamefont
  {Reid}(1968)}]{gage1968stability}%
  \BibitemOpen
  \bibfield  {author} {\bibinfo {author} {\bibfnamefont {K.}~\bibnamefont
  {Gage}}\ and\ \bibinfo {author} {\bibfnamefont {W.}~\bibnamefont {Reid}},\
  }\bibfield  {title} {\bibinfo {title} {The stability of thermally stratified
  plane poiseuille flow},\ }\href
  {https://www.cambridge.org/core/journals/journal-of-fluid-mechanics/article/stability-of-thermally-stratified-plane-poiseuille-flow/28045827147C938D28A95057ED6BADF2}
  {\bibfield  {journal} {\bibinfo  {journal} {Journal of Fluid Mechanics}\
  }\textbf {\bibinfo {volume} {33}},\ \bibinfo {pages} {21} (\bibinfo {year}
  {1968})}\BibitemShut {NoStop}%
\bibitem [{\citenamefont {Maslowe}\ and\ \citenamefont
  {Spiteri}(2013)}]{maslowe2013study}%
  \BibitemOpen
  \bibfield  {author} {\bibinfo {author} {\bibfnamefont {S.}~\bibnamefont
  {Maslowe}}\ and\ \bibinfo {author} {\bibfnamefont {R.}~\bibnamefont
  {Spiteri}},\ }\bibfield  {title} {\bibinfo {title} {A study of singular modes
  associated with over-reflection and related phenomena},\ }\href
  {https://www.cambridge.org/core/journals/journal-of-fluid-mechanics/article/abs/study-of-singular-modes-associated-with-overreflection-and-related-phenomena/0BD97053C34CC04FEF7418FDAD8A45B9}
  {\bibfield  {journal} {\bibinfo  {journal} {Journal of Fluid Mechanics}\
  }\textbf {\bibinfo {volume} {728}},\ \bibinfo {pages} {120} (\bibinfo {year}
  {2013})}\BibitemShut {NoStop}%
\bibitem [{\citenamefont {Kolokolov}\ and\ \citenamefont
  {Lebedev}(2016)}]{lebedev2016}%
  \BibitemOpen
  \bibfield  {author} {\bibinfo {author} {\bibfnamefont {I.~V.}\ \bibnamefont
  {Kolokolov}}\ and\ \bibinfo {author} {\bibfnamefont {V.~V.}\ \bibnamefont
  {Lebedev}},\ }\bibfield  {title} {\bibinfo {title} {Structure of coherent
  vortices generated by the inverse cascade of two-dimensional turbulence in a
  finite box},\ }\href {https://link.aps.org/doi/10.1103/PhysRevE.93.033104}
  {\bibfield  {journal} {\bibinfo  {journal} {Phys. Rev. E}\ }\textbf {\bibinfo
  {volume} {93}},\ \bibinfo {pages} {033104} (\bibinfo {year}
  {2016})}\BibitemShut {NoStop}%
\bibitem [{\citenamefont {Borovikov}(1990)}]{borovikov1990critical}%
  \BibitemOpen
  \bibfield  {author} {\bibinfo {author} {\bibfnamefont {V.}~\bibnamefont
  {Borovikov}},\ }\bibfield  {title} {\bibinfo {title} {Critical layer
  formation in a stratified medium with mean shear flows},\ }\href
  {https://link.springer.com/article/10.1007/BF01049822} {\bibfield  {journal}
  {\bibinfo  {journal} {Fluid Dynamics}\ }\textbf {\bibinfo {volume} {25}},\
  \bibinfo {pages} {397} (\bibinfo {year} {1990})}\BibitemShut {NoStop}%
\bibitem [{\citenamefont {Benney}\ and\ \citenamefont
  {Maslowe}(1975)}]{benney1975nlin}%
  \BibitemOpen
  \bibfield  {author} {\bibinfo {author} {\bibfnamefont {D.~J.}\ \bibnamefont
  {Benney}}\ and\ \bibinfo {author} {\bibfnamefont {S.~A.}\ \bibnamefont
  {Maslowe}},\ }\bibfield  {title} {\bibinfo {title} {The evolution in space
  and time of nonlinear waves in parallel shear flows},\ }\href
  {https://doi.org/10.1002/sapm1975543181} {\bibfield  {journal} {\bibinfo
  {journal} {Studies in Applied Mathematics}\ }\textbf {\bibinfo {volume}
  {54}},\ \bibinfo {pages} {181} (\bibinfo {year} {1975})}\BibitemShut
  {NoStop}%
\bibitem [{\citenamefont {Maslowe}(1986)}]{maslowe1986}%
  \BibitemOpen
  \bibfield  {author} {\bibinfo {author} {\bibfnamefont {S.~A.}\ \bibnamefont
  {Maslowe}},\ }\bibfield  {title} {\bibinfo {title} {Critical layers in shear
  flows},\ }\href
  {https://doi.org/https://doi.org/10.1146/annurev.fl.18.010186.002201}
  {\bibfield  {journal} {\bibinfo  {journal} {Annual Review of Fluid
  Mechanics}\ }\textbf {\bibinfo {volume} {18}},\ \bibinfo {pages} {405}
  (\bibinfo {year} {1986})}\BibitemShut {NoStop}%
\bibitem [{\citenamefont {Dörnbrack}\ and\ \citenamefont
  {Nappo}(1997)}]{dornbrack1997lincompare}%
  \BibitemOpen
  \bibfield  {author} {\bibinfo {author} {\bibfnamefont {A.}~\bibnamefont
  {Dörnbrack}}\ and\ \bibinfo {author} {\bibfnamefont {C.~J.}\ \bibnamefont
  {Nappo}},\ }\bibfield  {title} {\bibinfo {title} {A note on the application
  of linear wave theory at a critical level},\ }\href
  {https://doi.org/10.1023/A:1000270821161} {\bibfield  {journal} {\bibinfo
  {journal} {Boundary-Layer Meteorology}\ }\textbf {\bibinfo {volume} {82}},\
  \bibinfo {pages} {399} (\bibinfo {year} {1997})}\BibitemShut {NoStop}%
\bibitem [{\citenamefont {Thorpe}(1981)}]{thorpe1981experiment}%
  \BibitemOpen
  \bibfield  {author} {\bibinfo {author} {\bibfnamefont {S.~A.}\ \bibnamefont
  {Thorpe}},\ }\bibfield  {title} {\bibinfo {title} {An experimental study of
  critical layers},\ }\href {https://doi.org/10.1017/S0022112081001365}
  {\bibfield  {journal} {\bibinfo  {journal} {Journal of Fluid Mechanics}\
  }\textbf {\bibinfo {volume} {103}},\ \bibinfo {pages} {321–344} (\bibinfo
  {year} {1981})}\BibitemShut {NoStop}%
\bibitem [{\citenamefont {Campbell}\ and\ \citenamefont
  {Maslowe}(2003)}]{campbell2003}%
  \BibitemOpen
  \bibfield  {author} {\bibinfo {author} {\bibfnamefont {L.~J.}\ \bibnamefont
  {Campbell}}\ and\ \bibinfo {author} {\bibfnamefont {S.~A.}\ \bibnamefont
  {Maslowe}},\ }\bibfield  {title} {\bibinfo {title} {Nonlinear critical-layer
  evolution of a forced gravity wave packet},\ }\href@noop {} {\bibfield
  {journal} {\bibinfo  {journal} {Journal of Fluid Mechanics}\ }\textbf
  {\bibinfo {volume} {493}},\ \bibinfo {pages} {151–179} (\bibinfo {year}
  {2003})}\BibitemShut {NoStop}%
\bibitem [{\citenamefont {Landau}\ and\ \citenamefont
  {Lifshitz}(2013)}]{landau2013quantum}%
  \BibitemOpen
  \bibfield  {author} {\bibinfo {author} {\bibfnamefont {L.}~\bibnamefont
  {Landau}}\ and\ \bibinfo {author} {\bibfnamefont {E.}~\bibnamefont
  {Lifshitz}},\ }\href {https://books.google.ru/books?id=neBbAwAAQBAJ} {\emph
  {\bibinfo {title} {Quantum Mechanics: Non-Relativistic Theory}}},\ Course of
  theoretical physics, Vol. 3\ (\bibinfo  {publisher} {Pergamon},\ \bibinfo
  {year} {2013})\BibitemShut {NoStop}%
\end{thebibliography}%

\appendix

\section{Monochromatic wave equation in~cylindrical coordinates}		
\label{app.sec:cylindrical_eq}

In this Appendix we derive the equations for radial propagation of monochromatic wave in Fourier space $(\omega,r,\varphi,k_z)$ from the linear system (\ref{eq:lin-system_cyl}) for $\bm{u}$ dynamics in inviscid limit. Applying transformations of the derivatives in (\ref{homogeneity}), we introduce the relative wave frequency $\widetilde{\omega}(r)=\omega-m/rU(r)$ with Doppler shift from mean flow and $k_{\scriptscriptstyle}^2(r)=\left(m/r\right)^2+k_z^2$ that is square of perpendicular wavevector component, which arise in the following calculations. 

From $z$-component of system~(\ref{eq:lin-system_cyl}) we obtain the expression~(\ref{cyl:p-eq}) between vertical velocity $w$ and modified pressure $p$. The incompressibility condition
\[
	\frac{1}{r}\big(r v(r)\big)^\prime +\frac{im}{r}u+i k_z w=0
\] 
enables one to express them in terms of planar velocity components $u,v$:
\begin{equation}    \label{cyl:incmpr}
	p=\frac{\widetilde{\omega}}{k_{z}}w=\frac{i\widetilde{\omega}}{k_{z}^{2}}\left[\frac{1}{r}\big(rv\big)^{\prime}+\frac{im}{r}u\right].
\end{equation}

Next, we substitute in $r$-component of Eq.~(\ref{eq:lin-system_cyl}) curl,
\begin{equation*}
	\frac{m}{r}\widetilde{\omega}w-k_{z}\left[\widetilde{\omega}u+i\left(2\tilde{\Omega}(r)+\Sigma(r)\right)v\right]=0,
\end{equation*}
the formula for $w$ above, thus obtaining relation~(\ref{cyl:u-via-v}) between $u$ and $v$. Substitution of it in~(\ref{cyl:incmpr}) gives the remaining equation (\ref{cyl:w-eq}) of the system.

Then, we write $z$-component of Eq.~(\ref{eq:lin-system_cyl}) curl,
\[
	\frac{d}{dr}\left[r\left(\widetilde{\omega}u+(2\widetilde{\Omega}+\Sigma)iv\right)\right]-im\left(\widetilde{\omega}v-2i\widetilde{\Omega}u\right)=0,
\]
and substitute there expression~(\ref{cyl:u-via-v}) for $u$. Divided by $im$, this brings us to the closed ODE for component $v$:
\begin{multline}						\label{cyl:curl-z}
    \widetilde{\omega}\frac{d}{dr}\left(\frac{v^{\prime}+v/r}{k_{{\scriptscriptstyle \perp}}^{2}}\right)+rv\frac{d}{dr}\left[\frac{m}{r^{2}}\frac{2\widetilde{\Omega}+\Sigma}{k_{{\scriptscriptstyle \perp}}^{2}}\right]
	\\-\widetilde{\omega}v+\frac{2\widetilde{\Omega}\left(2\widetilde{\Omega}+\Sigma\right)k_{z}^{2}}{\widetilde{\omega}k_{{\scriptscriptstyle \perp}}^{2}}v=0,
\end{multline}
where we have used $\widetilde\Omega^\prime=-\Sigma/r$, $\widetilde\omega^\prime=-m\Sigma/r$, which is equivalent to Eq.~(\ref{eq:v-eq_cyl}) in the main text. 

To make a comparison with case of free inertial waves, we derive the closed ODE for modified pressure $p$: considering (\ref{cyl:incmpr}) along with $r$-component of (\ref{eq:lin-system_cyl}), one gets the relation
\begin{equation}						\label{eq:p-via-v}
	iv=\frac{2\tilde{\Omega}m/rp-\widetilde{\omega}p^{\prime}}{2\tilde{\Omega}\left(2\tilde{\Omega}+\Sigma\right)-\widetilde{\omega}^{2}}.
\end{equation}
By means of it we obtain from (\ref{cyl:incmpr}):
\begin{equation}						\label{eq:p-eq_cyl}
	\left[(2\tilde{\Omega}+\Sigma)\frac{m}{r}+\widetilde{\omega}\frac{d}{dr}\right]\frac{2\tilde{\Omega}mp-\widetilde{\omega}p^{\prime}r}{2\tilde{\Omega}\left(2\tilde{\Omega}+\Sigma\right)-\widetilde{\omega}^{2}}=rk_{{\scriptscriptstyle \perp}}^{2}p.
\end{equation}
In absence of the mean flow, $U\equiv 0$, frequencies $\widetilde \Omega,\widetilde \omega$ become constants, thereby bringing one to the Eq.~(\ref{eq:in-waves_bare}).

\section{Reflection of wave from the shear flow}		\label{app.sec:reflection}

Here we consider a monochromatic wave in the rectilinear geometry that reaches its critical layer from outside and study how its propagation in the mean flow $U(y)$ causes the reflected wave. Equation (\ref{eq:v_invisc}) can be written in the form 
\begin{align}           \label{refl:WKB-0}  
    & v^{\prime\prime} + k_y^2 v  = 0, 
    \\  \nonumber
    & k_{y}(y) =
    \bigg(\frac{2\Omega\big(2\Omega+\Sigma\big)k_{z}^{2}}{\widetilde{\omega}^{2}}-k_{{\scriptscriptstyle \perp}}^{2}-\frac{k_{x}\Sigma^{\prime}}{\widetilde{\omega}}
    \bigg)^{1/2}.
\end{align}

At $y\to-\infty$, where $U$ is absent, by choosing $v(y)$ in form of incident/reflected free wave, see~(\ref{eq:wave_asympt}), one discerns two different travelling wave behaviors (inwards and outwards from the mean flow, resp.) for solution of the equation. Similarly, near the critical layer $y\to y_\ast$ travelling wave solution takes one of the two approximate forms from (\ref{eq:zero_asympt}). Between these asymptotics, in case of general $U(y)$ profile, the global $v$ solution cannot be represented as superposition of travelling waves. Such waves can be distinguished in WKB-limit, though corrections to the approximation causes reflection from the mean flow~\cite{berry1972WKB}. So it should be possible to pose a scattering problem to investigate the wave propagation and reflection in terms of these asymptotics from both sides. However, the equation (\ref{refl:WKB-0}) should be transformed before that, as $k_y^2$ does not tends to zero at $y \to y_\ast$, so the standard one-dimensional scattering theory~\cite{landau2013quantum} cannot be applied.

To find the proper transformation, let us start from the WKB-solution. We introduce the phase based on positive WKB-wavenumber from~(\ref{refl:WKB-0}):
\begin{equation}				\label{refl:phase_coord}
    \phi = \intop^y_{y_0} d\xi \,k_y(\xi);\quad \frac{d}{k_{y}dy}\equiv \frac{d}{d\phi},
\end{equation}
In terms of $a$, see the text after (\ref{quasiclassic-sol}), equation (\ref{refl:WKB-0}) can be rewritten in the form 
\begin{equation}	\label{refl:WKB-1-1}
   \frac{d^2 a}{d\phi^2} + a
    = Va,
    \qquad
    V= \frac{1}{\sqrt{k_y}}\frac{d^2}{d\phi^2} \sqrt{k_y}.
\end{equation}
Although the potential $V$ goes to zero at $\phi\rightarrow-\infty$, it does not yet have the desired property, since it has constant limit at $\phi \to +\infty$ (i.e. at $y\to y_\ast$): $V\to \rho^2/4$. This issue can be resolved with the repeating of the procedure, which was applied to (\ref{refl:WKB-0}) to bring us to (\ref{refl:WKB-1-1}), now for the equation (\ref{refl:WKB-1-1}). We introduce corrected phase $\psi$ and amplitude $h$ and correction factor $q$ for the wavenumber,
\begin{equation}
    \psi = \intop_{0}^{\phi}q\, d\varphi=
    \intop^y_{y_0} d\xi \,qk_y,
    \quad 
    v = \frac{h}{\sqrt{q\,k_y}},
    \quad 
    q = \sqrt{1-V}.
\end{equation}
The initial equation (\ref{refl:WKB-0}) takes the form 
\begin{equation}\label{eq:03}
    \frac{d^2 h}{d\psi ^2}  + h = \mathcal{V}h, 
    \qquad 
    {\mathcal V} = \frac{1}{\sqrt{q}} \frac{d^2}{d\psi^2} \sqrt{q},
\end{equation}
where the new potential ${\mathcal V}(\psi)$ is localized around the point $\psi =0$ and tends to zero at $\psi\to\pm\infty$. 

We are now in position to utilize techniques from 1D QM scattering theory on a localized potential~\cite{landau2013quantum}. At $\psi \to \pm \infty$ a general solution of~(\ref{eq:03}) is $h=C_+ e^{i\psi}+ C_- e^{-i\psi}$ with constant $C_\pm$. At $\psi \rightarrow +\infty$, i.e. at $\eta\rightarrow -0$, the solution $v_\sigma$ takes the asymptotic of WKB-wave (\ref{eq:02}) transmitting through the layer:
\begin{align}                   \label{refl:inc_sol}
    h_{\sigma}
    =
    C_{-\sigma}e^{-i\sigma \psi}
\end{align}
i.e. 
\begin{align}                   \label{refl:asympts}
    v_{\sigma}
    =
    \frac{C_{-\sigma}e^{-i\sigma\psi}}{\sqrt{qk_{y}}}	
    =
    \frac{2C_{-\sigma}\left(-\eta\right)^{\frac{1+i\sigma\beta}{2}}}{\beta}e^{i\sigma \psi_{0}}.
\end{align}
Note that its asymptotic at $y\rightarrow -\infty$, $v_\sigma \sim e^{-i\sigma |k_y^{(0)}| y}$, agrees with inequality $\omega k_y^{(0)}<0$ for the incident wave. Below we choose $C_{-\sigma}=|k_y^{(0)}|^{1/2}$ at $y\to-\infty$ to make the incident wave of unity amplitude.

Intending to determine the amplitude $C_\sigma$ of reflected WKB-wave (with asymptotic $h_{-\sigma}$ at $\psi\rightarrow -\infty$), we write scattering theory equations. The Green function of a free WKB-wave problem in Fourier space is chosen as
\begin{equation}				\label{refl:Green-func}
	G_{\sigma}\left(\psi_{1},\psi_{2}\right)=\intop_{-\infty}^{+\infty}\frac{d\kappa}{2\pi}e^{i\kappa\left(\psi_{1}-\psi_{2}\right)}\frac{-1}{\kappa^{2}-1+0i\sigma},
\end{equation}
with asymptotic $G_\sigma \sim e^{i\sigma \psi}$ at $\psi\rightarrow -\infty$. While constructing perturbation theory on $\mathcal{V}$ for reflected wave $\chi_\sigma (\psi)$, we introduce scattering function $F_\sigma\left(\kappa,\kappa^\prime\right)$ to express the latter in Fourier space via the incident wave~$h^{(0)}_\sigma$:
\begin{equation}                \label{refl:F-func}
    \chi_\sigma(\kappa)=G_\sigma(\kappa)\int \frac{d\kappa^\prime}{2\pi}F_\sigma(\kappa,\kappa^\prime)h_\sigma^{(0)}(\kappa^\prime).
\end{equation}
In this way, the reflection amplitude can be formulated
\[
	r=\frac{i \sigma |k_y^{(0)}|^{1/2}}{2}F_\sigma(\sigma,-\sigma)
\]
via the $F_\sigma$ function that satisfies the equation:
\begin{equation}				\label{refl:F-eqn}
	F(\kappa,\kappa^\prime)=\mathcal{V}(\kappa-\kappa^\prime)+\int\frac{d\mathit{p}}{2\pi}\mathcal{V}(\kappa-\mathit{p})G(\mathit{p})F(\mathit{p},\kappa^\prime).
\end{equation}

In the limit~(\ref{quasiclassic-cond}) of small perturbation to WKB approximation one can consider $\mathcal{V}$ as such and apply the Born approximation for scattering function $F_{\sigma}(\kappa,\kappa^{\prime})\approx\mathcal{V}\left(\kappa-\kappa^{\prime}\right)$. In particular, one should keep the leading contribution produced by the small quantity $V$ (\ref{refl:WKB-1-1}). Then one obtains the explicit formula for $r$:
\begin{equation}                \label{refl:Born-r_coeff}
    r=
    \frac{i \sigma |k_y^{(0)}|^{1/2}}{2}\int d\psi\, 
    e^{2 i \sigma\psi}\mathcal{V}.
\end{equation}
Finding approximate expression for the potential ${\mathcal V}$ as
\begin{multline}\label{refl:Born_approx}
    \mathcal{V}(y)
    \approx  
    -
    \frac{1}{4}\left(\frac{d}{k_{y}dy}\right)^{\!2}
    \frac{1}{\sqrt{k_y}}
    \left(\frac{d}{k_{y}dy}\right)^{\!2}
    \sqrt{k_y}
    ,
\end{multline} 
we arrive to Eq.~(\ref{eq:reflection}) in the main text after integration by parts in (\ref{refl:Born-r_coeff}).

\section{Derivation of equations with viscous terms}	\label{app.sec:viscous_eqs}

Here we first derive the Eqs.~(\ref{visc:u_eq},\ref{visc:v_eq}) for rectilinear model~(\ref{eq:linear_wave}) with a finite viscosity in Fourier space. Starting with its curl:
\begin{gather}				\label{veq:wx}
		\left(\widetilde{\omega}-i\nu\Delta\right)\varpi_{x}-i\varpi_{y}U^{\prime}-k_{z}uU^{\prime}=-2\Omega k_{z}u,
		\\					\label{veq:wy}
		\left(\widetilde{\omega}-i\nu\Delta\right)\varpi_{y}-k_{z}vU^{\prime}=-2\Omega k_{z}v,
		\\					\label{veq:wz}				
		\left(\widetilde{\omega}-i\nu\Delta\right)\varpi_{z}-ivU^{\prime\prime}=-\left(2\Omega-U^{\prime}\right)k_{z}w,
\end{gather}
where $\bm \varpi = \mathrm{curl} \:\bm v$ is vorticity vector and $\Delta=\partial_y^2-k_{\scriptscriptstyle\perp}^2$ is Laplace operator, we consider first its $x$ and $z$-components and substitute an expression~(\ref{eq:uz}) for $w$ there. In order to exclude viscous term with $u$ we take their linear combination $k_z$(\ref{veq:wx})$-k_x$(\ref{veq:wz}), i.e.~y-component of Eq.~(\ref{eq:linear_wave})'s~$\mathrm{curl}\:\mathrm{curl}$:
\begin{equation}				\label{veq:curl}
	\left(\widetilde{\omega}-i\nu\Delta\right)\Delta v-2i\Omega k_{{\scriptscriptstyle \perp}}^{2}u-2\Omega k_{x}v^{\prime}+k_{x}U^{\prime\prime}v=0,
\end{equation}
that gives us expression $u$ via $v$ component in Eq.~(\ref{visc:u_eq}). 
Taking~(\ref{veq:wy}) multiplied by $k_z$,
\begin{equation}
	\left(\widetilde{\omega}-i\nu\Delta\right)\left(ik_{{\scriptscriptstyle \perp}}^{2}u+k_{x}v^{\prime}\right)+\left(2\Omega-U^{\prime}\right)k_{z}^{2}v=0,
\end{equation} 
with substitution~(\ref{visc:u_eq}) one arrives to closed equation for $v(y)$. After multiplication by $2\Omega$ we obtain~(\ref{visc:v_eq}).

Next, we derive a solutions for the viscous equation in the vicinity of the critical layer. The solution $v_\pm$ of (\ref{visc:v_pm-eq}) can be obtained with the Laplace method for differential equations. In the notations from the Appendix A2 of book~\cite{drazin_reid_book2004} it is proportional to the linear combination of functions $A_k(z,p)$:
\begin{gather}
    A_1(z,p)+A_2(z,p)+A_3(z,p)\propto 
    \intop_{L}d\chi\,\frac{e^{-i\sigma\eta\chi-(\eta_\mathrm{v}\chi)^{3}/3}}{\chi^{\left(3\pm i\beta\right)/2}},    \nonumber
    \\
    z= -i\sigma\eta/\eta_\mathrm{v},\qquad p=(3\pm i\beta)/2.
    \label{visc:v-sol_Airy}
\end{gather}
In (\ref{visc:v-sol_Airy}) there is contour integral in $\chi$-complex plane cut along the semi-axis $\left(0,+\infty\right)$. The contour of integration $L$ begins and ends at infinity $\Re\, \chi\rightarrow +\infty$; it passes round the branch point $\chi=0$ to the other side from the branch cut. In order make the singularity $\chi=0$ integrable, we apply integration by parts, obtaining derivative of the exponent on the way. After that, the contour integral can be replaced by integral along the semi-axis since difference between integrand values on each side of the branch cut is proportional to the one itself. Thus, we obtain integral representation (\ref{visc:v-sol_real}) on the real axis.

\end{document}